\documentclass[1p]{elsarticle}
\usepackage[latin1]{inputenc}
\usepackage{amsmath}
\usepackage{amsfonts}
\usepackage{amssymb}
\usepackage{graphicx}
\usepackage{mathrsfs}
\usepackage{xcolor}

\let\abs=\envert

\newcommand\ket[1]{\big|\,#1\,\big\rangle}
\newcommand\bra[1]{\big\langle\,#1\,\big|}

\newcommand\op[1]{{\bf #1}}  
\newcommand\opc[1]{{\cal #1}}
\newcommand\vect[1]{\vec{\op{#1}}}    

\newcommand\ic{{\rm i}}
\newcommand\un[1]{{\rm\,#1}}
\newcommand\unc[1]{\,[{\rm #1}]}

\newcommand\trs[2]{\mathrm{Tr}_{#1}\!\left\{#2\right\}}
\newcommand\trbs[2]{\mathrm{Tr}_{#1}\!\big\{#2\big\}}

\newcommand\numeq[1]{(\ref{#1})}
\newcommand\eq[1]{Eq.~(\ref{#1})}

\newcommand\fig[1]{Fig.~\ref{#1}}
\newcommand\numfig[1]{\ref{#1}}
\newcommand\figs[1]{Figs.~\ref{#1}}
\newcommand\sect[1]{Section~\ref{#1}}

\newcommand\apen[1]{\ref{#1}}

\newcommand\bib[1]{\cite{#1}}
\newcommand\tabla[1]{Table~\ref{#1}}

\def \H {{\cal{H}}}
\def \intef {\int_{-{\infty}}^{\infty}}

\newcommand\insertfig[6]
{\begin{figure}[#1]
    \begin{center}
        \includegraphics[bb = 0 0 #4cm #5cm]{#3.jpg}
    \end{center}
    \caption{#6}
    \label{#2}
\end{figure}}

\begin{document}

\begin{frontmatter}

    \title{Evolution towards quasi-equilibrium in nematic liquid crystals studied through decoherence of multi-spin multiple-quantum coherences}
    \author{H. H. Segnorile\corref{cor1}}
    \ead{hector.segnorile@unc.edu.ar}
    \author{C. E. Gonz\'alez}
    \ead{ceciliae.gonzalez@unc.edu.ar}
    \author{R. C. Zamar}
    \ead{ricardo.zamar@unc.edu.ar}
    \cortext[cor1]{Corresponding author}
    \address{Instituto de F\'isica Enrique Gaviola - CONICET,
    Facultad de Matem\'atica, Astronom\'{\i}a y F\'{\i}sica,
    Universidad Nacional de C\'ordoba \\ M.Allende y H. de la Torre - Ciudad Universitaria,
    X5016LAE - C\'ordoba, Argentina}

    \begin{abstract}
        New evidence is presented in favor of irreversible decoherence as the mechanism which leads an initial out-of-equilibrium state to quasi-equilibrium in nematic liquid crystals. The NMR experiment combines the Jeener-Broekaert sequence with reversal of the dipolar evolution and decoding of multiple-quantum coherences to allow visualizing the evolution of the multi-spin coherence spectra during the formation of the quasi-equilibrium states.
        We vary the reversion strategies and the preparation of initial states and observe that the spectra amplitude attenuate with the reversion time, and notably, that the decay is frequency selective.
        We interpret this effect as evidence of ``eigen-selection'',  a signature of the occurrence of irreversible adiabatic decoherence, which indicates that the spin system in liquid crystal NMR experiments conforms an actual open quantum system.
    \end{abstract}


\end{frontmatter}


\section{Introduction}\label{se:Intro}

    The quantum dynamics that drives a quantum system of interacting particles from an initial coherent state into a thermodynamic state is still a matter of concern both in fundamental questions and practical applications. Nuclear spins  are privileged systems whose quantum dynamics develops over a wide range of timescales that can be accurately probed through solid state Nuclear Magnetic Resonance (NMR).	
    During their irreversible transit from an initial coherent state to the final thermal equilibrium, spins in solids and complex fluids may go through intermediate ``quasi-equilibrium'' (QE) states that evolve slowly towards thermal equilibrium due to spin-lattice relaxation only. The build-up of QE in solids and liquid crystals (LC) takes place within an intermediate time scale, shorter than relaxation but longer than the typical free induction decay (FID).
    The possibility of preparing different initial quantum states and manipulating and witnessing their dynamics turns nuclear spins into excellent test-beds for studying the microscopic mechanisms underlying the build-up of QE.		
    This work aims to provide further experimental evidence on its origin.  Particularly, we study the spin states prepared with the phase-shifted Jeener-Broekaert pulse pair (JB). The preparation time (time between pulses) determines the number of correlated spins and also the kind of QE state into which the system evolves after the waiting time.

    We consider the system formed by the proton spins in liquid crystals. In this kind of sample, each molecule constitutes a small cluster of interacting spins, coupled with a highly correlated non-spin environment.  In fact, due to the rapid individual molecular motions, the intermolecular spin interactions average out; still, a strong intramolecular dipolar energy remains because of the high degree of long range molecular orientation typical of the mesophase \cite{degen93,shyki88,dongbook}.
    The NMR spectroscopic properties of LC are similar to those of solids; that is, the on-resonance FID signal evolves over a short timescale determined by the local dipolar couplings within the molecule, and the spin-lattice relaxation timescale is much longer than that of the FID.
    The $^1$H NMR spectra in nematic LC generally show partially resolved doublets that reflect the anisotropic spin distribution within LC molecules.
    Likewise, despite the small number of relevant spin degrees of freedom, the LC spin clusters can also develop QE states within a timescale similar to solids \cite{pintar74,schmiedel82}. Indeed, the experimental results clearly showed that it is possible to selectively prepare several QE states in LC by adequately setting the preparation pulse sequences \cite{MenGzzZam05,SegBonGzzAcoZam09,Bonin13}.

    Owing to the small size of the spin system, the occurrence of QE in LC can not be explained through arguments borrowed from statistical mechanics, valid for large interacting systems, like spin diffusion  \cite{pintar74}. In previous works, we anticipated that the development of QE should instead be related to irreversible adiabatic decoherence \cite{SegZam11,SegZam13}. This mechanism, which operates in the first stage of the quantum evolution, is associated with the spin system coupling with a large correlated quantum environment.
    Such environment-induced decoherence without dissipation erodes the off-diagonal elements of the spin density matrix in the ``preferred'' basis \cite{Zurek03}, driving the system to a state with a diagonal-in-block form. A synthetic description of this phenomenon is given in \sect{se:decoherence}.

    Experiments on NMR signal refocusing \cite{SegZam13,GzzSegZam11} showed that  irreversible decoherence and development of QE take place within the same characteristic timescale. 
    Further theoretical and experimental work \cite{SegZam11,SegZam13,GzzSegZam11,SegGzzZam_Annals21} indicated that a small cluster of interacting spins, viewed as an open quantum system coupled to a correlated environment, can indeed undergo irreversible adiabatic decoherence, just as the experimentally observed in LC.
    A significant feature predicted by the theoretical proposal is that decoherence is controlled by the eigenfrequencies of the spin part of the {\em interaction} Hamiltonian rather than those of the system Hamiltonian. This phenomenon implies a kind of {\em selection} (induced by the coupling with the environment) since each element of the density operator decays with a characteristic time, scaled by the corresponding eigenvalue differences, and was accordingly called  ``eigen-selection'' \cite{footnote_eigensel}.
    This effect was experimentally observed on the single-spin single-quantum spectra of some nematic LCs \cite{SegZam13}. The experiment showed that the higher frequency components of the reverted signal spectra attenuate faster than the lower frequency components, and this behavior could not be attributed to experimental non-idealities (non-reverted terms, pulse setting errors, etc.).

    A similar phenomenon might be expected in the many-spin coherences that develop after the JB pulse pair, during the build-up of QE. In this work, we study the evolution of the many-spin coherences with an NMR experiment designed to visualize both their coherence amplitude and spectra. A frequency-selective attenuation of such spectra can thus indicate that LC also behave as open quantum systems within this stage.
    The technique combines the JB preparation \cite{jeener67} with reversion of the dynamics and multiple quantum coherences (MQC) encoding, with the aim of displaying the effect of eigen-selection (attenuation and frequency compression on the spectra) during decoherence. We are particularly interested in exploring the transient to QE states other than the standard dipolar order (two-spin order), featured by a higher multi-spin character. We expect to detect the traces of decoherence in the non-equilibrium many-spin coherence spectra in several nematic LC samples.

    The small-sized spin systems of LC allows for exact numerical calculation of the unitary quantum dynamics under the spin Hamiltonian (isolated system). In fact, considering the dipolar couplings within a single LC molecule suffice to describe the main characteristics of NMR signals, which enables comparison with the experiment.

    In \sect{se:QEDetect} we explain general aspects of the experiment which tracks the evolution of coherences and spectra during the buildup of QE states and contains an analytical description of the signals. \sect{se:QED_ExpSetMeas} shows the experimental results, and in \sect{se:Conclu} we discuss the results.
    \apen{ap:non-ideal} is added to show the effect of non-idealities on the outcome of the experiment on a hypothetically closed system.

	\subsection{Background}

	\subsubsection{On the time scales of the spin dynamics} \label{se:timescale}
    The experimental phenomenology on thermotropic LC \cite{Bonin13,GzzSegZam11,bulju09} shows that different time scales are manifest in the spin dynamics. Their occurrence is determined by the system-environment coupling nature.
		
	When the observed spin system in an NMR experiment is considered as an open quantum system, the total Hamiltonian is written as
	\begin{equation}\label{eq:def_H}
	   \opc{H} = \opc{H}_S + \opc{H}_E + \opc{H}_{SE},
	\end{equation}
    where the system of interest is described, in isolation, by $\opc{H}_S$;  $\opc{H}_E$ is the Hamiltonian of the environment and  $\opc{H}_{SE}$ represents the system-environment interaction which has both spin and non-spin variables (e.g. interaction between spins and mechanical variables, or a boson bath, etc.). When the description is made in the rotating frame (at the Larmor frequency) the main contribution to the system Hamiltonian is the (secular -- high field) dipole-dipole interaction, i.e., $\H_S \equiv \H_D$.

    The system-environment Hamiltonian can be written as a linear combination of product operators with the general form
    \begin{equation}\label{eq:def_HSE}
        \opc{H}_{SE} = \sum_q \op{A}^{q(s)} \otimes \op{F}^{\,q(e)},
    \end{equation}
    where $\op{A}^{q(s)}$ acts on the spin variables and $\op{F}^{\,q(e)}$ on the environment variables. The superscripts $(s)$ and $(e)$ stress  the Hilbert space (system $S$ or environment $E$) where the operator belongs. The general form of \eq{eq:def_HSE} may be particularized to different kinds of materials, such as nematic LC or an ordered array of spin 1/2 pairs in a phonon environment with strong intra-pair dipolar interactions. In the case of LC the index $q$ labels molecules and $\op{A}^{q(s)} \equiv \H_{Dq}^{(s)}/S_{zz}$, where $\H_{Dq}^{(s)}$ is the secular dipolar Hamiltonian of the $q$-th molecule (including all the intra-molecular interactions). The operator
    $\op{F}^{\,q(e)} \cong \op{S}_{\op{zz}q}^{(e)}-S_{zz}\op{1}^{(e)}$, where $\op{S}_{\op{zz}q}^{(e)}$ is the quantum operator that represents the `$\op{zz}$' order parameter tensor component of the $q$-th molecule (its eigenvalues depend on the long molecular axis polar angle orientation) and $S_{zz}$ is its equilibrium value across the whole sample.
    In the solid-like example of \cite{DomGzzSegZam16} the index $q$ labels spin pairs and $\op{A}^{q(s)}$ is proportional to the intra-pair secular dipolar Hamiltonian of the $q$-th pair, the environment operator is defined as $\op{F}^{\,q(e)} \equiv \sum_{\bf k} \left(g^{q*}_{\bf k}\,\op{b}^{(e)}_{\bf k} + g^{q}_{\bf k}\,\op{b}^{\dagger(e)}_{\bf k}\right)$, where index $\bf{k}$ stands for wavenumber vectors $\vec{k}$ and normal mode $l$, $\op{b}^{\dagger(e)}_{\bf k}$ and $\op{b}^{(e)}_{\bf k}$ are respectively the $\bf{k}$-phonon creation and annihilation operators, and $g^{q}_{\bf k}$ is the coupling strength constant of the $q$-th pair.

    From a theoretical viewpoint, commutations between the intervening Hamiltonians, which can be assumed with a good degree of approximation at different stages of the dynamics, allow us to understand the different time scales.
    A first distinction emerges from $ \left[\opc{H}_S,\opc{H}_{SE} \right] $.  The adiabatic regime is characterized by
    \begin{equation}\label{eq:adiaba}
        \left[\opc{H}_S,\opc{H}_{SE}\right] = 0.
    \end{equation}
    which implies that $\left\langle \opc{H}_S\right\rangle $ is a constant of motion.  More precisely, condition \numeq{eq:adiaba} describes an \emph{essentially adiabatic system} (see Section II.B in \cite{SegZam11}). Let us emphasize that adiabaticity does not impose a restriction over the model system but only on the time scale where it can be assumed. It is similar to the high-field approximation in NMR, which neglects the nonsecular terms of the dipolar Hamiltonian contribution to the time evolution within a bounded range.
    Contrarily, the condition $ \left[\opc{H}_S,\opc{H}_{SE} \right] \neq 0  $ 	identifies the long term regime, where the energy-exchange affects the spin dynamics.
    This is the time scale of relaxation or thermalization where the observed system approaches thermal equilibrium with the environment.

    Within the adiabatic condition, another property worth consideration concerns the environment and system-environment Hamiltonians commutator. Since generally the Hamiltonian norms satisfy $\| \opc{H}_E \| > \| \opc{H}_S \|$, the time range where $[\opc{H}_E,\opc{H}_{SE}] \sim 0$ is valid may be shorter than the one where  $[\opc{H}_S,\opc{H}_{SE}] \sim 0$.  Within this period the system is \emph{essentialy isolated}
    and the environment variables can be treated as time independent functions (not operators). This is the regime where the FID signals are observable and attenuate mainly due to interference (typically a few hundreds of microseconds in LC), even though coherences may not decay irreversibly.
    A detailed analysis of the relation between time scales and commutation relations  can be found in Refs.\cite{SegZam11,SegZam13,DomGzzSegZam16}.	
    There is still another stage in the system dynamics, where the commutation of the environment and system-environment Hamiltonians cannot be assumed,
    namely $\left[\opc{H}_E,\opc{H}_{SE} \right] \neq 0$. In this work we are interested in this regime, where the quantum character of the environment operators is crucial and conditions the spin dynamics (see section \ref{se:decoherence}), and coherences attenuate irreversibly without energy exchange with the environment. This intermediate time scale,  shorter than relaxation but longer than the FIDs is typically less than 10 ms in LC.

    Besides the assumptions on the commutation properties of  $\opc{H}_S,\opc{H}_{SE}$ and $\opc{H}_E$, it is usual to qualify the system-environment coupling according to the relative celerity of the system evolution and the environment fluctuations. Assuming a Markovian coupling of a quantum open system implies considering that a time scale exists where the observed system does not disturb the environment modes, which remain thermalized during the whole evolution \cite{Shresta_etal,Privman03}. This assumption (coarse graining) allows describing the system dynamics in terms of a quantum master equation of the Lindblad (or GKSL) form \cite{GorKossSud76,Lindblad76,BreuerPetruccione}, as in the usual NMR relaxation theory \cite{abragam61,Redfield65}, where the diagonal elements of the density matrix approach thermal equilibrium at temperature ${\rm T}$ with a characteristic spin-lattice relaxation time $T_1$. The adiabatic limit of the Markovian theory describes the secular contribution to the off-diagonal density-matrix elements attenuation characterized by a decay time $T_2$ (this $T_2$ differs from the inverse of the linewidth).

    It is an experimental fact that coherences evolve in times much less than spin-lattice relaxation times $T_1$ or $T_{1D}$ (by several orders in many cases).  Indeed, it was experimentally shown \cite{GzzSegZam11} that in 8CB (4-cyano-4'-$n$-octyl-H$_{11}$-biphenyl) the decoherence time in the nematic phase is about 20 times shorter than $T_2$ measured in the isotropic phase \cite{footnote_T2}. Thus, it is not adequate to apply a coarse-grain description to explain a phenomenon occurring within an early scale, dictated by the correlation between the many-interacting-spins system and the environment modes.

    The fact that $T_1$ process is a slow mechanism compared with the characteristic time scale of the NMR signals frequently leads to treating  the intermediate time scale dynamics as that of a genuinely closed system that undergoes a unitary evolution \cite{PRB_Cory-etal,multispBoutis12,SkrebZaripov_99}. However, there is no physical reason for disregarding the rapid and strong correlation (without energy exchange) between the system and environment degrees of freedom \cite{schlosshauer_19} within this time scale. A non-Markovian theoretical approach that describes the environment reduced density operator without the coarse-grain assumption is summarized in Section \ref{se:decoherence}.

    The terminology used in the literature is not homogeneous about the usage of the terms ``decoherence'' and ``relaxation''. Decoherence may allude to different phenomena: some work use it to refer the secular contribution to $T_2$, that is, to coherence loss with no change of the energy level populations \cite{abragam61}. In the field of quantum optics and quantum information this phenomenon is the result of ``quantum noise'' \cite{Palma96}.
    The same term may also indicate the decay of the MQ coherence NMR signals in spin clusters \cite{Kroj-Suter_04,Kroj-Suter_06,Bochkin18}. In this work, we refer as relaxation to the spin-lattice interaction mechanisms ($T_1$ or $T_{1D}$ processes) that bring the spin system to thermal equilibrium with the environment. On the other hand, we refer to quantum decoherence in the context of the spin-boson model \cite{BreuerPetruccione}, to indicate the decay of the off-diagonal elements of the density matrix (on the eigenbasis of the spin-environment Hamiltonian) due to the adiabatic coupling of the observed quantum system with a correlated quantum environment, as explained in the next section.

    \subsubsection{Outline of decoherence in an open quantum system} \label{se:decoherence}

    We summarize a description of \emph{adiabatic quantum decoherence} (AQD) of an observed system in contact with an external unobserved bath.
    Since we are interested in describing processes which take place in a time scale earlier than relaxation and thermalization, we assume that the Hamiltonians in \eqref{eq:def_H} satisfy the \emph{adiabatic} condition \numeq{eq:adiaba}.
    This implies that $\opc{H}_S$ and the spin part of $\opc{H}_{SE}$ have a common eigenbasis $\{\ket{m}\}$. Physically, it means excluding system-environment energy exchange and allows factorizing the total evolution operator as
    \begin{equation}\label{eq:opevol_T}
        \op{U}_T(t) = e^{-\ic\,\opc{H}_{S}\,t} e^{-\ic\,(\opc{H}_{SE}+\opc{H}_E)\,t} \equiv \op{U}_S(t)  \op{U}_{SE}(t),
    \end{equation}
    where
    \begin{equation}\label{eq:Op_free}
        \op{U}_S(t) \equiv e^{-\ic\,\opc{H}_{S}\,t}
    \end{equation}
    acts on the spin degrees of freedom only and describes the dynamics of a closed system. The quantum \emph{openness} enters in the evolution under the non-commuting terms $\opc{H}_{SE}+\opc{H}_E$
    \begin{equation}\label{eq:Op_SL}
        \op{U}_{SE}(t) \equiv e^{-\ic\,(\opc{H}_{SE}+\opc{H}_E)\,t}.
    \end{equation}
    When $\op U_T(t)$ acts on a separable initial state of the compound system, it correlates the $S$- and $E$-degrees of freedom and the state $\rho(t)$ is no longer separable (see Section II.B from \cite{SegZam11} or Section II.A from \cite{SegZam13} for details about the dynamics under these operators).

    The expectation value of any operator acting on the observable $S$-space can be written in terms of the density operator reduced over the $E$-space, defined as
    \begin{equation}\label{eq:sigma_t}
        \sigma(t) \equiv \trs{E}{\rho(t)} =  \trs{E}{\op{U}_T(t)\,\rho(0)\,\op{U}_T(t)^\dagger},
    \end{equation}
    where the partial trace $\trs{E}{\cdot}$ runs over the environment variables. It is worth to note that the time evolution of $\sigma$ can be non-unitary.
    A separable initial state represented by a factorized density operator $\rho(0)= \rho_S(0) \rho_E(0)$ (with $\rho_S(0)$ and $\rho_E(0)$ the system and environment equilibrium states respectively) evolves into a state, whose matrix elements in the common basis $\{\ket{m}\}$ have the general form \cite{BreuerPetruccione,Yukalov11}
    \begin{equation}\label{eq:sigma_mn}
        \sigma_{mn}(t) = \sigma_{mn}^{isol}(t)\,G_{mn}(t).
    \end{equation}
    The first factor corresponds to the evolution of the system in isolation (uncoupled to the environment), under the spin Hamiltonian only
    \begin{equation}\label{eq:sigma_isol}
        \sigma_{mn}^{isol}(t)= \bra{m} \op{U}_S(t)\,\rho_S(0)\,\op{U}_S^{\dagger}(t)\ket{n}.
    \end{equation}
    The second factor, on the contrary, contains the coupling with the environment and is called \emph{adiabatic quantum decoherence function}
    \begin{equation}\label{eq:G_mn_tr}
        G_{mn}(t) = \trs{E}{\op{U}_E^\dagger(m,t)\,\op{U}_E(n,t)\,\rho_E(0)},
    \end{equation}
    where $\op{U}_E(n,t)$ is a unitary operator acting on $E$-space, and is defined by
    \begin{equation}\label{eq:U_E}
        \op{U}_{SE}\ket{n} = \ket{n}\op{U}_E(n,t),
    \end{equation}
    namely, it is the result of evaluating the spin part of the interaction Hamiltonian in $\op{U}_{SE}$ on its $n$-th eigenvalue \cite{BreuerPetruccione}.
    From \numeq{eq:G_mn_tr}, $G_{m,n}=1$ for $m=n$, which means that the diagonal matrix elements are unaffected by decoherence. In the limit of a large environment, the decoherence function decreases irreversibly. Schematically,
    \begin{equation}\label{eq:G_mn}
        G_{mn}(t) \equiv e^{\Gamma_{mn}(t)}\;,
    \end{equation}
    where $\Gamma_{mn}(t)$ is a complex function, whose real part is a negative function that diverges for $t \rightarrow \infty$. The structure of   $\Gamma_{mn}(t)$ depends on the properties of the non-commuting Hamiltonians $\opc H_E$ and $\opc H_{SE}$, which dictate the time scale of decoherence.
    This function was calculated for different cases, as the spin-boson model in ensambles of uncorrelated units \cite{Palma96,privman98,Luczka90}, correlated spin clusters in nematic LC molecules \cite{SegZam11,SegZam13}, and spin-pairs coupled with a correlated phonon bath \cite{DomGzzSegZam16}.
    Particularly, the decoherence function corresponding to both free-evolution and time reversal experiments in nematic LC was calculated in Refs.\cite{SegZam11,SegZam13}, where the dynamics is governed by the Hamiltonians described after \eq{eq:def_HSE}.
    In that case, the dynamics of a single molecule is representative of the whole the sample, but each molecule is affected by all the rest through the quantum correlation introduced by the coupling with the bath.
    Therefore, $\{\ket{m}\}$ is the eigenbasis of the secular dipolar Hamiltonian of one molecule, where
    the reduced density matrix \numeq{eq:sigma_t} and the decoherence function \numeq{eq:G_mn_tr} are expressions for a single molecule containing correlations with the whole sample.
    The corresponding decoherence function can be written as
    \begin{equation}\label{eq:G_mn_LC}
        G_{mn}(t) \simeq e^{-\abs{\lambda_m -\lambda_n}^{p}\,\hat{\Gamma}_{\Delta S}(t)}\,
    \prod_j e^{-\abs{\lambda_m -\lambda_n}^{r_j}\,\hat{\Gamma}_{C_j}(t)}\;,
    \end{equation}
    where $\hat{\Gamma}$ are positive real functions satisfying $\lim_{\,t \rightarrow \infty}\hat{\Gamma} = +\infty$, $\{p,r_j\} \in \mathbb{N}^+$, and $\lambda_m$ is the eigenvalue of the spin part of the interaction Hamiltonian \numeq{eq:def_HSE} for the $q$-th molecule,
    $\op{A}^{q(s)}\ket{m} = (\H_{Dq}^{(s)}/S_{zz})\ket{m} = \lambda_m\ket{m}$.
    The function $\hat{\Gamma}_{\Delta S}(t)$ and the exponent $p$ depend on the  eigenvalue distribution of the operator $\op{S}_{\op{zz}q}^{(e)}-S_{zz}\op{1}^{(e)}$, i.e., it depends on the orientational molecular distribution function (OMDF), and the characteristic decay time of this exponential factor has the same order as the FID signal.
    On the other hand, the functions $\hat{\Gamma}_{C_j}(t)$ and the exponents $r_j$ depend on the eigenvalue distribution of nesting commutators involving operators $\op{F}^{\,q(e)}$ of different molecules and these operators with the environment Hamiltonian $\opc{H}_E^{(e)}$, such commutators are indexed by $j$ in \numeq{eq:G_mn_LC}.

    We can see in \eq{eq:G_mn_LC} that the decaying exponential factors depend on the eigenvalue difference $\abs{\lambda_m -\lambda_n}$, then, larger  differences imply faster decay; besides, the matrix elements \numeq{eq:sigma_mn} conecting equal eigenvalues $\lambda_m$ will be preserved from decoherence. Such kind of selective effect was called \emph{eigen-selectivity} \cite{SegZam11,SegZam13} and is an ubiquitous characteristic of AQD, which is implicit in the general case \numeq{eq:G_mn} and was also observed in the case of solid spin systems coupled to quantum environments \cite{DomGzzSegZam16,privman98}.
    Accordingly, $\sigma(t)$ attains a diagonal-in-blocks structure (in the preferred basis $\{\ket{m}\}$) after a characteristic decoherence time.
    Such a state  becomes time independent (quasi-equilibrium) and may only evolve when the adiabatic condition \numeq{eq:adiaba} ceases to hold, that is, when spin-lattice relaxation becomes relevant within a longer time scale.

    Deducing an analytical expression for functions $\Gamma (t)$ can be a hard task, depending on the form of the spin, environment, and interaction Hamiltonians of the particular model. In a favorable case (analytically speaking) like an array of spin pairs coupled to a phonon environment \cite{SegGzzZam_Annals21,DomZamSegGzz17} it is possible to evaluate a decoherence function that allows quantitative comparison with the experiment. Otherwise, a model for nematic LC like the one outlined below \eq{eq:def_HSE} allows writing an expression for $\Gamma_{\Delta S}$ (see \eq{eq:Gmn_exp_free}) however it does not yield an analytical expression for ${\Gamma}_{C_j}(t)$. Fortunately, their main features can be deduced without their explicit functional form. For example, ref.\cite{SegZam11} shows that the characteristic decay time of the exponential factors with $\hat{\Gamma}_{C_j}$ is longer than the decay time due to $\hat{\Gamma}_{\Delta S}$; thus, within the early times where the FID is observable it is safe to approximate
    \begin{equation}\label{eq:G_mn_LC_aproxfree}
        G_{mn}(t) \simeq e^{-\abs{\lambda_m -\lambda_n}^{p}\,\hat{\Gamma}_{\Delta S}(t)}\;.
    \end{equation}
    An important characteristic of the functions $\hat{\Gamma}_{C_j}$ is that they cancel if $\left[\opc{H}_E,\opc{H}_{SE}\right] = 0$, leaving \numeq{eq:G_mn_LC_aproxfree} as the only decoherence factor in \eq{eq:G_mn_LC}. Such a condition corresponds to an {\it essentially isolated} system, which undergoes decoherence and is decidedly different to considering a closed system.
    If one instead disregard the quantum character of the environment variables in the interaction Hamiltonian \numeq{eq:def_HSE}, which amounts to assuming a closed system description, decoherence has no effect at all.
    In that case, the NMR signals measured in LC with few spins in a molecule would not vanish in a short time scale, which contradicts the experiment.
    Alternatively (without using the AQD scheme), one might deduce a decaying dynamics within a closed-system description by introducing an \emph{ad hoc} statistical distribution of the geometric variables. However, such unitary evolution should, in principle, be reversed and the apparent attenuation of observed signals (either single- or multiple-quantum) recovered. On the contrary, the reversion experiments show that the initial coherent states can in fact only partially be recovered and that coherences extinguish over a time scale larger than that of the FID by one or a few orders of magnitude, but still much shorter than $T_1$. This irreversible dynamics is evidenced in reversion experiments in LC \cite{SegZam13,GzzSegZam11} and also in solids \cite{DomZamSegGzz17}.
    Particularly, Ref.\cite{SegZam11} shows that in LC it is possible, in principle, to refine reversion pulse sequences in order to attain  $\hat{\Gamma}_{\Delta S}(t) = 0\;(\forall\,t)$ and to compensate some of the terms $\hat{\Gamma}_{C_j}(t)$ but not cancelling all the $\hat{\Gamma}$ functions.
    That is the concept or vision of irreversibility which we ascribe in this work.
    The usual reversion experiments, like those used in this work, allow cancelling the short time decoherence described by  $\hat{\Gamma}_{\Delta S}$ so we can write the decoherence function under reversion  $(rt)$ as
    \begin{equation}\label{eq:G_mn_LC_rt}
        G_{mn}^{\,(rt)}(t) \simeq \prod_j e^{-\abs{\lambda_m -\lambda_n}^{r_j}\,\hat{\Gamma}_{C_j}^{\,(rt)}(t)}\;.
    \end{equation}

    Under this vision, decoherence and relaxation are physical phenomena corresponding to different time limits of the dynamics, and both emerge from a full quantum description of the interaction Hamiltonian \numeq{eq:def_HSE}.
    Interestingly, in the NMR bibliography, it is natural to deal with the environment as a quantum entity within relaxation theories. In contrast, coherences are usually treated in the frame of  closed systems, where decays obey reversible interference (or dephasing), ruling out irreversible decoherence.
    As commented above, the open quantum system approach allows describing both the early time scale, characterized by fast decay of the FID signals, and also the eigen-selective irreversible attenuation, that can be observed along the intermediate time scale in reversion experiments.
    The effect of AQD is to bring an initial state into one whose density matrix is diagonal in blocks in the eigenbasis of the interaction Hamiltonian. Since this is actually the structure of quasi-equilibrium states, and also QE states establish along the same intermediate time scale, in this work we intend to provide experimental evidence to test AQD as the mechanism behind the occurrence of quasi-equilibrium.

\section{Description of the pulse sequence}\label{se:QEDetect}

    The experiment was tailored to study the way in which decoherence affects the frequency content of the spectra of different coherences through eigen-selection, during the transit from a coherent multi-spin initial state to the QE.
    The pulse sequence shown in \fig{fig:Puls}(a) combines the Jeener-Broekaert (JB) experiment with encoding of multiple-quantum coherences and refocalization of the spin dynamics under the secular dipolar Hamiltonian.
    The standard JB  sequence [$(\pi/2)_x - t_p - (\pi/4)_y - t - (\pi/4)_y$] (top row in \fig{fig:Puls}(a), with $\tau=0$ and $\varphi = 0^\circ$) is traditionally used to prepare a QE state sometimes called `dipolar order'. The first $(\pi/2)_x$ pulse produces  an initial one-spin single-quantum coherence state which evolves freely under the secular dipolar Hamiltonian during the preparation time $t_p$, developing  multi-spin, single-quantum coherences. The second pulse $(\pi/4)_y$ rotates this state into multi-spin multiple-quantum coherences (zero, single, double or higher quantum order). The state so prepared evolves freely during a waiting time $t$.
    A last read pulse $(\pi/4)_{y}$ projects part of the state into observable single-quantum coherence. If $t$ is `long enough' the observed signal is consistent with a diagonal-in-block density operator (on the basis of the dipolar Hamiltonian) and its amplitude only depends on $t$ in the scale of relaxation. Let us call $t_{QE}$ to the timescale needed for the particular spin system to reach QE.	
    Experiments show that the signal shape after the read pulse depends on the preparation time, in fact, at least three different kinds of QE states were observed in nematic liquid crystals \cite{Bonin13}.

    \insertfig{ht}{fig:Puls}{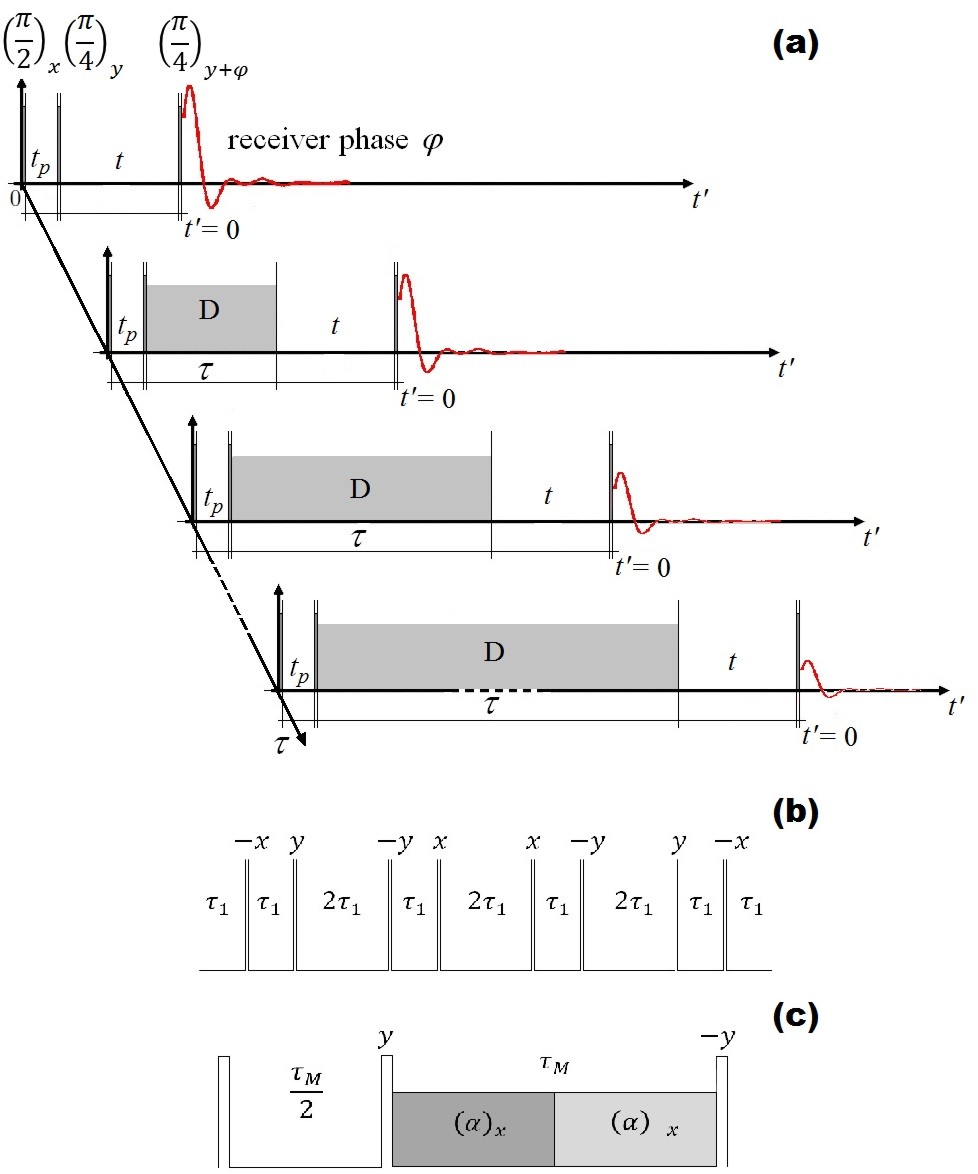}{9.2}{11}{(a) The pulse sequence used to trace the formation of quasi-equililibrim starts with the phase-shifted pulse pair of the JB sequence, where $t_p$ allows selecting the QE state that develops at the end of the sequence.  The waiting time $t$ and the phase $\varphi$ of the read pulse are increased in steps. Fourier Transform on $t$ and $\varphi$ gives the spectra of each coherence order separately. The reversion block $D$ of length $\tau$, suspends evolution under the secular dipolar Hamiltonian, highlighting the effect of decoherence while evolving towards QE. The reversion sequences within $D$ blocks are (b): the MREV8, where $\tau = 12\,\tau_1$; or (c): `magic-sandwich' , where $\tau= 1.5\,\tau_M$}

	The sequence depicted in rows 2 to 4 of \fig{fig:Puls}(a) aims to make visible the effect of decoherence during the development of QE,
    we thus
    \begin{itemize}
    	\item  prepare a coherent state with the phase-shifted pulse pair of the JB sequence.
    	\item use the phase $\varphi$ of the read pulse to encode the different coherence orders \cite{cho03},
        \item repeat the experiment while incrementing the waiting time $t$ in steps so that a Fourier Transform on $t$ and $\varphi$ gives the spectra of each coherence order separately \cite{footnote_faseprop}.
    	\item scrutinize the evolution during the build-up of QE $\tau < t_{QE}$, by increasing length of block $D$.
    \end{itemize}

    Block $D$ probes the important information on the effect of decoherence over the amplitudes and coherence spectra. The action of each $D$ block is to suspend the free-evolution under the secular dipolar Hamiltonian during  $\tau$. Its effect is to `undo' the reversible evolution and thus disclose the irreversible effect of decoherence. The consequences of the irreversible mechanisms on the signal become observable by incrementing the duration of block $D$.
    We use two different schemes for the reversed dynamics to ensure that the results are independent of the particular experimental strategy. That is, block $D$ may either be the MREV8 sequence \cite{SegZam13,mansf71,RhimEV73:58,RhimEV73:59}, schematized in \fig{fig:Puls}(b), or the `magic-sandwich' (MS) sequence \cite{GzzSegZam11,RhimPW70,RhimPW71,RhimK71,RhimK72,nielsen97}, described in \fig{fig:Puls}(c).
    Each strategy offers particular advantages: MREV8 can be both used with a continuous variation of the time $\tau_1$ or by concatenating blocks of this sequence \cite{SegZam13}, allowing a thorough scan of the decoherence time. MS involves less (although longer) pulses, and is consequently less vulnerable to errors in the pulse settings.
    After the $D$ block there is a free-evolution period driven by $\op{U}_S(t)\op{U}_{SE}(t)$. By varying its length $t$ in $N_t$ experiments the observed signal encodes the frequency content given by the evolution  $\op{U}_S(t)$, under the dipole-dipole Hamiltonian.

    In summary, \fig{fig:Puls}(a) illustrates the whole experiment, where a set of sequences at different refocusing times $\tau$ are vertically arranged  in order to imitate the two-dimensional disposition chosen to show the experimental results (see \sect{se:QED_ExpSetMeas}).
    Each row represents the set of $N_t \times N_{\varphi}$ experiments  needed to calculate the spectra of all the encoded coherences (where  $N_{\varphi}$ the number of phase steps in the read pulse).

\subsection{Analytic description}\label{se:AnDes}

    Let us now write the output signals of the proposed experiment and the interpretation of the corresponding spectra.
    The initial equilibrium state of the complete system can be written as the tensor product
    $\rho_{(eq)}= \rho_{S(eq)}\rho_{E(eq)}$, where $\rho_{S(eq)} \propto \op{I}_{\op{z}}$ (we omit the identity operator because it does not contribute to the signals in NMR experiments) represents the spin system in equilibrium with an external field $\vect{B_0}$, in the high temperature regime, and $\rho_{E(eq)}$ is the density operator of the non-spin degrees of freedom (which is not affected by the RF pulses nor by the evolution under only-spin operators).

    After the first  $(\pi/2)_x$ pulse, the obtained state $\rho_S(0)$ evolves under the total evolution operator of \eq{eq:opevol_T}. However, it is worth to notice that both the preparation time $t_p$ and the acquisition time $t'$, used in the experiment depicted in \fig{fig:Puls}, are much shorter (few tens of microseconds) than the timescale scanned by the sum $\tau + t$ (tens to hundreds of microseconds).
    Experimental evidence \cite{SegZam11,GzzSegZam11} shows that the dynamics driven by $\op{U}_{SE}$ is slower than that driven by $\op{U}_S$ so we can consider
    \begin{equation}\label{eq:short_t_approx}
        \op{U}_{SE}(t) \sim \op{1}\;\text{during short times $t$},
    \end{equation}
    which justifies assuming a purely unitary dynamics during $t_p$ and $t'$.
    Then, under the assumption \numeq{eq:short_t_approx} the state at the end of the preparation period is
    \begin{equation}\label{eq:rho_tp}
        \rho(t_p) = \op{U}_S(t_p)\,\rho_S(0)\,\op{U}_S^{\dag}(t_p) \: \rho_{E(eq)} = \rho_S(t_p)\,\rho_{E(eq)},
    \end{equation}
    with $\rho_S(t_p) \equiv \op{U}_S(t_p)\,\rho_S(0)\,\op{U}_S^{\dag}(t_p)$.\\
    On the contrary, \emph{we expect the effect of decoherence, due to the inevitable quantum coupling with the environment, to manifest during $\tau + t$.}

    The second pulse  $(\pi/4)_y$ transforms $\rho(t_p)$ in a density operator with zero, single, double or higher quantum order coherence, so that immediatly after preparation
    \begin{equation}\label{eq:rho_tpp}
        \rho(t_p^+) = \op{R}_{\op{y}}(\pi/4)\,\rho_S(t_p)\,\op{R}_{\op{y}}(-\pi/4) \: \rho_{E(eq)} = \rho_S(t_p^+)\,\rho_{E(eq)},
    \end{equation}
	where $\rho_S(t_p^+) \equiv \op{R}_{\op{y}}(\pi/4)\,\rho_S(t_p)\,\op{R}_{\op{y}}(-\pi/4)$ and the pulse operator
    $\op{R}_{\alpha}(\theta) \equiv e^{\,\ic\,\op{I}_{\alpha}\theta}$.
	
    During the interval $\tau$  the state is driven through a forward-backwards evolution, which has the effect of suspending the dynamics under the secular dipolar Hamiltonian, but, in principle, cannot undo evolution under $\opc{H}_{SE}+\opc{H}_E$. Let us represent the evolution within block $D$ by the operator $\op{U}_D(\tau)$ \cite{footnote_oprev}.
    It is important to note that the spin part of the effective Hamiltonian in both reversion strategies used in our work, MREV8 (\fig{fig:Puls}(b)) and MS (\fig{fig:Puls}(c)), is proportional to the secular dipolar Hamiltonian \cite{SegZam11,SegZam13}. This means that during $\tau$ the system evolves under the combination of the free-evolution Hamiltonian \numeq{eq:def_H} and the reversion Hamiltonian
    $\opc{H}_{(rt)} = \opc{H}_{S(rt)} + \opc{H}_E + \opc{H}_{SE(rt)}$, where $\opc{H}_{S(rt)} = -\opc{H}_{D}/2$ and $\opc{H}_{SE(rt)}$ has the same general form as \eq{eq:def_HSE} but $\op{A}^{q(s)} = -\opc{H}_{Dq}^{(s)}/(2\,S_{zz})$.  In this way, the adiabatic condition \numeq{eq:adiaba} is also accomplished under the reversion dynamics ($\left[\opc{H}_{S(rt)},\opc{H}_{SE(rt)}\right] = 0$) and the spin part of the free-evolution cancels with that of the reversion Hamiltonian.
    In \apen{ap:non-ideal}, we study the effect of non-idealities in the particular setting of the MREV8 sequence. However, using a different strategy as the MS sequence allows us to compare different reversion dynamics and to detect if an experimental artifact is present.\\
    Therefore, we write the state after the reversion period ($\tau$) and the free-evolution (no reversion during $t$) as
    \begin{equation}\label{eq:rho_t}
    \begin{split}
        \rho(t,\tau,t_p) = \op{U}_T(t)\,\op{U}_D(\tau)\,\rho_S(t_p^+)\rho_{E(eq)}\op{U}_D^{\dag}(\tau)\,\op{U}^{\dag}_T(t)\:,
    \end{split}
    \end{equation}
    with $\op{U}_T(t)$ defined in \eq{eq:opevol_T}.

    The read pulse $\op{R}_{\op{y}+\varphi}(\pi/4)$ projects the multiple-quantum coherences over a measurable single-quantum coherence.
    Its phase $\pi/2+\varphi$ (and consequently the receiver phase) is systematically varied to encode the coherence orders.

    Since the NMR signals are expectation values of spin operators we use the density operator reduced over the environment \numeq{eq:sigma_t}; this is,
    $\sigma(t,\tau,t_p) \equiv \trs{E}{\rho(t,\tau,t_p)}$.
    Then, the signal $S_{\alpha}$ (with $\alpha = \op{x}$ or $\op{y}$), as a function of the acquisition time $t'$ and the  other variable parameters of the experiment ($\varphi, t, \tau$ and $t_p$), is
    \begin{equation}\label{eq:S_def1}
    \begin{split}
        &S_{\alpha}(t',\varphi,t,\tau,t_p) = \trbs{S}{\op{I}_{\alpha}\,\op{U}_S(t')\,\op{R}_{\op{y}+\varphi}(\pi/4)\\
        &\qquad\qquad\qquad\times \sigma(t,\tau,t_p)\,\op{R}_{\op{y}+\varphi}(-\pi/4)\,\op{U}_S^{\dag}(t')},
    \end{split}
    \end{equation}
    where we also used the approximation \numeq{eq:short_t_approx}.
    In \eq{eq:S_def1}, the operators inside the trace apply only over the system or spin Hilbert space and must to have the superscript $(s)$, in the following we omit this superscript for simplicity.
    Using that
    \begin{subequations}\label{eq:Op_def}
    \begin{equation}\label{eq:Op_def1}
        \op{R}_{\op{y}+\varphi}(\pi/4) = \op{R}_{\op{z}}(-\varphi)\,\op{R}_{\op{y}}(\pi/4)\,\op{R}_{\op{z}}(\varphi),
    \end{equation}
    \begin{equation}\label{eq:Op_def2}
        \op{I}_{(\op{x},\op{y})+\varphi} = \op{R}_{\op{z}}(-\varphi)\,\op{I}_{(\op{x},\op{y})}\,\op{R}_{\op{z}}(\varphi),
    \end{equation}
    \end{subequations}
    the signal can be written as
    \begin{equation}\label{eq:S_def2}
    \begin{split}
        S_{\alpha} &= \trbs{S}{\op{I}_{\alpha}\,\op{U}_S(t')\,\op{R}_{\op{y}}(\pi/4)\,\op{R}_{\op{z}}(\varphi)\,\sigma(t,\tau,t_p)\\
        &\qquad\qquad\qquad\times\op{R}_{\op{z}}(-\varphi)\,\op{R}_{\op{y}}(-\pi/4)\,\op{U}_S^{\dag}(t')}.
    \end{split}
    \end{equation}

    In order to disclose the coherence encoding implicit in \eq{eq:S_def2}, it is useful to write the reduced density operator in terms of a set of irreducible tensors $\{\op{T}^\Lambda_{\zeta \eta}\}$  which serve as an operator basis for the Liouville space where $\sigma$ belongs. Index $\zeta$ is the tensor rank,  $\eta$ the component number, and $\Lambda$ distinguishes between different species of tensors with the same rank \cite{weitek83}.
    The reduced density operator at any time is formally
    \begin{equation}\label{eq:sigma_tgen}
        \sigma(t,\tau,t_p) = \sum_{\Lambda}\sum_{\zeta \eta} \xi^\Lambda_{\zeta \eta}(t,\tau,t_p)\,\op{T}^\Lambda_{\zeta \eta},
    \end{equation}
    where
    \begin{equation}\label{eq:chi}
        \xi^\Lambda_{\zeta \eta}(t,\tau,t_p) = \trbs{S}{\op{T}^{\dag\Lambda}_{\zeta \eta}\,\sigma(t,\tau,t_p)}
    \end{equation}
    is the projection of $\sigma(t)$ on $\op{T}^{\dag\Lambda}_{\zeta \eta}$, with the orthonormal relation
    \[\trbs{S}{\op{T}^{\dag\Lambda}_{\zeta \eta}\,\op{T}^{\Lambda'}_{\zeta' \eta'}} = \delta_{\Lambda,\Lambda'}\delta_{\zeta,\zeta'}\delta_{\eta,\eta'},\]
    and $\delta$ is the Kronecker delta.
    It  is worth to notice that  $\xi^\Lambda_{\zeta \eta}$ as a function of $t$ (with fixed $t_p$ and $\tau$)  characterizes the tensor content variation of the density operator while evolving under the complete evolution operator $\op U_T(t)$.

    The reduced density operator $\sigma(t,\tau,t_p)$ in \eqref{eq:sigma_tgen} already encompasses the effect of the reversion block. Within the AQD approach of \sect{se:decoherence} its general expression in nematic LC is \cite{SegZam13}
    \begin{equation}\label{eq:sigma_deco}
        \sigma_{mn}(t,\tau,t_p) = \sigma_{mn}^{isol}(t,t_p)\,G_{mn}(t)\,G_{mn}^{\,(rt)}(\tau),
    \end{equation}
    where
    \begin{equation}\label{eq:sigma_isol_exp}
        \sigma_{mn}^{isol}(t,t_p) = \sigma_{mn}^{\,0}(t_p)\,e^{-\ic\,\left(\lambda_m -\lambda_n\right)\,S_{zz}\,t}
    \end{equation}
    are the density matrix elements evolved under the spin Hamiltonian only, and do not depend on $\tau$.
    In \eq{eq:sigma_isol_exp} we define $\sigma^{\,0}(t_p) \equiv \sigma(0,0,t_p) = \rho_S^{(s)}(t_p^+)$.\\
    The effect of  coupling with the environment comes in the functions $G_{mn}(t)$ (mainly essentially isolated decoherence) and $G_{mn}^{\,(rt)}(\tau)$ (irreversible decoherence) which are respectively driven by $\hat{\Gamma}_{\Delta S}$ and $\hat{\Gamma}_{C_j}$. When particularizing to nematic LCs, $G_{mn}(t)$ takes the form \cite{SegZam13}
    \begin{equation}\label{eq:Gmn_exp_free}
    \begin{split}
        G_{mn}(t) &\simeq \int_{-\infty}^{\infty} d(\Delta S)\,e^{\,\ic\,2\pi\nu_{mn}\,t\,\Delta S/S_{zz}}\,p(\Delta S)\\
                    &= 2\pi\,\hat{p}\left(\frac{2\pi\nu_{mn}}{S_{zz}}\,t\right),
    \end{split}
    \end{equation}
    where we defined the dipolar interaction frequencies $\nu_{mn} \equiv \frac{1}{2\pi}\left(\lambda_n -\lambda_m\right) S_{zz}$ in $\un{Hz}$ units,
    \[\hat{p}(\hat{\omega}) \equiv \opc{F}^{-1}_{\Delta S\rightarrow \hat{\omega}}\{p(\Delta S)\}(\hat{\omega})\] is the inverse Fourier transform of $p$,
    with $p(\Delta S)$ the orientational molecular distribution function (OMDF) associated with the eigenvalue distribution of the operator $\op{S}_{\op{zz}q}^{(e)}-S_{zz}\op{1}^{(e)}$.

   In this way, the main features of the complete dynamics in nematic LCs under decoherence can be summarized as:
    \begin{itemize}
        \item The characteristic decay time of the essentially isolated regime is similar to that of the FID. Decoherence within this regime is reversible.
        \item Decoherence driven by $\hat{\Gamma}_{C_j}$ emerges in a longer time scale and is irreversible.
        \item both regimes present eigen-selectivity.
    \end{itemize}

    In order to visualize how decoherence affects our experiment we use expression \numeq{eq:sigma_deco} in \eq{eq:chi} to obtain
    \begin{equation}\label{eq:chi_deco}
    \begin{split}
        \xi^\Lambda_{\zeta \eta}(t,\tau,t_p) &= 2\pi\,\sum_{mn} \bra{n}\op{T}^{\dag\Lambda}_{\zeta \eta}\ket{m}\,
        \sigma_{mn}^{\,0}(t_p)\,e^{\,\ic\,2\pi\nu_{mn}\,t}\\
        &\qquad\qquad\times\hat{p}\left(\frac{2\pi\nu_{mn}}{S_{zz}}\,t\right)\,G_{mn}^{\,(rt)}(\tau).
    \end{split}
    \end{equation}
    Finally, using expansion \numeq{eq:sigma_tgen} in \eq{eq:S_def2}, and the fact that
    \begin{equation}\label{eq:prop_TRz}
        \op{R}_{\op{z}}(\varphi)\,\op{T}^\Lambda_{\zeta \eta}\,\op{R}_{\op{z}}(-\varphi) = e^{\,\ic\,\eta\,\varphi}\,\op{T}^\Lambda_{\zeta \eta},
    \end{equation}
    the NMR signals measured under the sequences of \fig{fig:Puls} can be expressed as
    \begin{equation}\label{eq:S_ge}
    \begin{split}
        S_{\alpha}(t',\varphi,t,\tau,t_p) = \! \sum_{\eta} e^{\,\ic\,\eta\,\varphi} \sum_{\Lambda\,\zeta} g^{\Lambda,\alpha}_{\zeta \eta}(t')\,
        \xi^\Lambda_{\zeta \eta}(t,\tau,t_p),
    \end{split}
    \end{equation}
    where
    \begin{equation}\label{eq:func_g}
    \begin{split}
        g^{\Lambda,\alpha}_{\zeta \eta}(t') &\equiv \trbs{S}{\op{I}_{\alpha}
        \op{U}_S(t')\op{R}_{\op{y}}(\pi/4)\op{T}^\Lambda_{\zeta \eta}\\
        &\qquad\qquad\qquad\quad\times\op{R}_{\op{y}}(-\pi/4)\op{U}_S^{\dag}(t')}.
    \end{split}
    \end{equation}
    In our experiment, instead of $g^{\Lambda,\alpha}_{\zeta \eta}(t')$  we used $g^{\Lambda,\alpha}_{\zeta \eta}({t'_m})$, the average value of the signals within an interval centered at the first absolute maximum.

    The spectra of coherences emerge after calculating the Fourier transform of \eq{eq:S_ge} over the variables $t$ and $\varphi$, that is
    \begin{equation}\label{eq:S_Fourier_ge}
    \begin{split}
	    &\opc{F}_{\varphi\rightarrow\mu,\,t\rightarrow\nu}\left\{S_{\alpha}\right\}(t'_m,\mu,\nu,\tau,t_p)\\
	    &\qquad\qquad\qquad\qquad= \sum_{\eta} \delta(\mu-\eta) f_{\eta}(t'_m,\nu,\tau,t_p),
    \end{split}
    \end{equation}
    where $\mu$ is the conjugate variable of $\varphi$, $\eta$ labels the coherence number, and $\nu$ is the conjugate variable of time $t$.
    The function $f_{\eta}(t'_m,\nu,\tau,t_p)$ represents the spectrum of coherence order $\eta$ for a given combination of preparation time $t_p$ and evolution under reversed dynamics in $\tau$ (averaged around the measurement time $t'_m$), and is given by
    \begin{equation}\label{eq:nq_cspec}
    \begin{split}
        &f_{\mu}(t'_m,\nu,\tau,t_p)\\
        &\qquad\quad\equiv \sum_{\Lambda\,\zeta} g^{\Lambda,\alpha}_{\zeta, \mu}(t'_m)\,
        \opc{F}_{t\rightarrow\nu}\left\{\xi^{\Lambda}_{\zeta, \mu}\right\}(\nu,\tau,t_p).
    \end{split}
    \end{equation}
    The Fourier transform in \numeq{eq:nq_cspec} can be written using \eq{eq:chi_deco}. First, we observe that
    \begin{equation}\label{eq:FT_sAQD}
    \begin{split}
        &\opc{F}_{t \rightarrow \omega}\left\{e^{\,\ic\,2\pi\nu_{mn}\,t}\,\hat{p}\left(\frac{2\pi\nu_{mn}}{S_{zz}}\,t\right)\right\}(\omega)\\
        &= \opc{F}_{t \rightarrow \omega}\left\{\hat{p}\left(\frac{2\pi\nu_{mn}}{S_{zz}}\,t\right)\right\}\left(2\pi(\nu - \nu_{mn})\right)\\
        &= \frac{S_{zz}}{2\pi\abs{\nu_{mn}}}\,p\left(\frac{S_{zz}}{\nu_{mn}}\,(\nu - \nu_{mn})\right),
    \end{split}
    \end{equation}
    where we used the translation and scaling property of the Fourier transform.
    Then, we use \numeq{eq:FT_sAQD} to obtain
    \begin{equation}\label{eq:FT_xi}
    \begin{split}
        &\opc{F}_{t\rightarrow\nu}\left\{\xi^{\Lambda}_{\zeta, \mu}\right\}(\nu,\tau,t_p) =
        \sum_{mn} \bra{m}\op{T}^{\Lambda}_{\zeta \mu}\ket{n}^*\,\sigma_{mn}^{\,0}(t_p)\\
        &\qquad\times \frac{S_{zz}}{\abs{\nu_{mn}}}\,p\left(\frac{S_{zz}}{\nu_{mn}}\,(\nu - \nu_{mn})\right)\,G_{mn}^{\,(rt)}(\tau),
    \end{split}
    \end{equation}
    where $\bra{n}\op{T}^{\dag\Lambda}_{\zeta \mu}\ket{m} = \bra{m}\op{T}^{\Lambda}_{\zeta \mu}\ket{n}^*$. In this way \eq{eq:FT_xi} represents a spectrum defined by the dipolar frequencies $\nu_{mn}$, where each spectral component is determined by  the orientational molecular distribution
    function $p$ and scaled by the frequency $\nu_{mn}$. This means that larger $\nu_{mn}$ contribute with broader and smaller OMDF.
    Then, we can see that AQD determines the spectra shapes.
    Besides this, each frequency component amplitude is modulated in the time $\tau$ by the irreversible AQD, $G_{mn}^{\,(rt)}(\tau)$, which is eigen-selective; thus, frequency components centered at larger $\nu_{mn}$ are subject to faster decay.
    This connection, between the frequency line of the spectrum (obtained under the free-evolution dynamics in $t$) and the eigen-selectivity (produced by the irreversible AQD in $\tau$), is a consequence of the adiabatic condition \numeq{eq:adiaba}, and also of the fact that the interaction Hamiltonian under reversion and free-evolution are proportional, which produces the same dependence with the eigenstates of the preferred basis for different decoherence functions (as shown the \eq{eq:sigma_deco}).
    It is worth noting that AQD depends on the eigenvalues of the dipolar Hamiltonian, due to the definition \numeq{eq:def_HSE} of $\H_{SE}$, and it is independent of the eigenvalues of the Zeeman Hamiltonian; thus, the eigen-selection only refers to dipolar frequencies, and is independent (in principle) of the coherence order, as experimentally probed in ref. \cite{GzzSegZam11}.
    In conclusion, the total spectrum \numeq{eq:S_Fourier_ge} is the combination of all the coherence spectra, and each the frequency components of coherence order $\mu$ spectrum is affected by AQD; besides, every component is scaled by the matrix elements of the initial condition $\sigma^{\,0}(t_p)$, the tensor component $\op{T}^{\Lambda}_{\zeta \mu}$, and the factor $g^{\Lambda,\alpha}_{\zeta \eta}({t'_m})$.\\

    It is worth to note, the novelty that \eq{eq:nq_cspec} introduces comes in the non-unitary dynamics along the intermediate timescale $\tau + t$, which enters in the observed spectra through the Fourier transform of the coefficient $\xi(t,\tau,t_p)$. Particularly, the evolution under $\op U_S$ during $t$ endows each coherence signal with the frequency content of the dipolar Hamiltonian, which becomes evident in the corresponding spectra.
    If the observed system were actually closed, its NMR signal would not depend on $\tau$ because the dynamics would be completely reversible. However, such a reversible behaviour could be a fictitious expectation since actual physical systems are not closed.
    Thus, any observed  dependence on $\tau$ might be interpreted either as an indication of an irreversible (non-unitary) behaviour like the one outlined in \sect{se:decoherence}, or as the result of an incorrect experimental setting.\\

    As a final comment, it is worth pointing out that the AQD approach outlined here is developed under the high-temperature assumption, which, in NMR, holds for temperatures ${\rm T} > 10\un{K}$. This working hypothesis is adequate for treating LC since the typical temperature of mesophases is in the order of room temperature. In the case of a solid-like array of weakly interacting spin pairs, the sensibility of decoherence with temperature is negligible at room temperature \cite{DomZamSegGzz17}.
    Recent work \cite{SegGzzZam_Annals21} showed that a solid ensemble of noninteracting spin-pairs coupled with a correlated phonon environment exhibits an efficient temperature-independent AQD mechanism. Accordingly, AQD effects may be expected at all temperatures, which, in turn, enables the occurrence of QE states even within the $\mu K$ range. This implies the possibility of studying quantum correlations (such as entanglement or discord) by using dipolar QE states, for example, under low Zeeman and high dipolar temperature conditions, as in Ref.\cite{LazarevFeld20}, which can have applications in the field of quantum computing or quantum information.\\

    In the next section, we compare the features of coherence spectral evolution under the decoherence and the relaxation process.

\subsection{Comparison of features between decoherence and relaxation processes}\label{se:CompDecRelax}

    In this section, we study the effects of a relaxation process on the coherence dynamics. We aim to compare the characteristic decay times and the effects on the spectra of the signals, or the density matrix, predicted by a relaxation theory regarding the decoherence process studied in \sect{se:decoherence}.
    In order of that, we write the high-temperature form of the general Markovian master equation, which represents the spin dynamics in the `coarse-grained' time scale, is \cite{abragam61,Redfield65}
    \begin{equation}\label{eq:EcMaestra}
    \frac{d\sigma(\tau)}{d\tau} = -\frac{1}{2} \intef d\tau' \,\trs{E}{[ \opc{H}_{SE}(\tau),\; [ \opc{H}_{SE}(\tau-\tau'),\, \Delta\sigma(\tau)]]\,\rho_E},
    \end{equation}
    where $\Delta\sigma(\tau) \equiv \sigma(\tau) - \sigma_{eq}$ and  $\sigma_{eq}$ is the equilibrium density matrix.
    The system-environment Hamiltonian can be developed as following
    \begin{equation}\label{eq:devHSE}
    \H_{SE}(\tau) = \sum_q \sum_{i<j} \op{A}^q_{ij}(\tau)\,\Delta\op{F}^{\,q}_{ij}(\tau),
    \end{equation}
    with $\Delta\op{F}^{\,q}_{ij}(\tau) \equiv \op{F}^{\,q}_{ij}(\tau) - \overline{\op{F}^{\,q}_{ij}}$ and
    \[\op{A}^q_{ij}(\tau) \cong e^{\,\ic\opc{H}_Z\,\tau} \op{A}^q_{ij} e^{-\ic\opc{H}_Z\,\tau} = \op{A}^q_{ij}\,e^{-\ic\,q\,\omega_0\,\tau}.\]
    Using the Markovian approximation, we can write
    \begin{equation}\label{eq:devHSE_Markov}
    \begin{split}
    \H_{SE}(\tau) &= \sum_q \sum_{i<j} \op{A}^q_{ij}\,\Delta\op{F}^{\,q}_{ij}\,e^{-\ic\,q\,\omega_0\,\tau},\\
    \H_{SE}(\tau-\tau') &= \sum_q \sum_{i<j} \op{A}^q_{ij}\,\Delta\op{F}^{\,q}_{ij}(-\tau')\,e^{-\ic\,q\,\omega_0\,(\tau-\tau')}.
    \end{split}
    \end{equation}
    Then, under the secular approximation we can replace $e^{-\ic\,(q-q')\,\omega_0\,\tau} \rightarrow \delta_{q,q'}$, and \eq{eq:EcMaestra} can be developed as
    \begin{equation}\label{eq:EcMaestra_devJ}
    \begin{split}
    \frac{d\sigma(\tau)}{d\tau} &= -\frac{1}{2} \sum_q \sum_{i<j,i'<j'} \bigg\{
    \op{A}^{-q}_{i'j'}\,\op{A}^{q}_{ij}\,\Delta\sigma(\tau)\,J^{(I)\,q}_{i'j',ij}(\omega_0)
    - \op{A}^{q}_{ij}\,\Delta\sigma(\tau)\,\op{A}^{-q}_{i'j'}\,J^{(II)\,q}_{i'j',ij}(\omega_0)\\
    &\qquad\qquad\quad + \Delta\sigma(\tau)\,\op{A}^{q}_{ij}\,\op{A}^{-q}_{i'j'}\,J^{(II)\,q}_{i'j',ij}(\omega_0)
    - \op{A}^{-q}_{i'j'}\,\Delta\sigma(\tau)\,\op{A}^{q}_{ij}\,J^{(I)\,q}_{i'j',ij}(\omega_0) \bigg\},
    \end{split}
    \end{equation}
    where we defined
    \begin{subequations}\label{eq:Def_J}
    \begin{equation}\label{eq:Def_JI}
        J^{(I)\,q}_{i'j',ij}(\omega) \equiv \intef d\tau\;e^{-\ic\,q\,\omega\,\tau}\,\trs{E}{\Delta\op{F}^{-q}_{i'j'}\,\Delta\op{F}^{\,q}_{ij}(-\tau)\,\rho_E},
    \end{equation}
    \begin{equation}\label{eq:Def_JII}
        J^{(II)\,q}_{i'j',ij}(\omega) \equiv \intef d\tau\;e^{-\ic\,q\,\omega\,\tau}\,\trs{E}{\Delta\op{F}^{\,q}_{ij}(-\tau)\,\Delta\op{F}^{-q}_{i'j'}\,\rho_E}.
    \end{equation}
    \end{subequations}
    The correlation functions \numeq{eq:Def_J} accomplish the following relationship
    \begin{equation}\label{eq:Rel_JI_JII}
        J^{(II)\,q}_{i'j',ij}(\omega) = e^{-\beta_0\,\omega}\,J^{(I)\,q}_{i'j',ij}(\omega),
    \end{equation}
    where $\beta_0 \equiv \hbar/(K_B\,T)$, with $K_B$ is the Boltzmann constant and $T$ is the absolute temperature.
    Therefore, in the high-temperature approach we have
     \begin{equation}\label{eq:Rel_JI_JII_HT}
        J^{(II)\,q}_{i'j',ij}(\omega) \simeq J^{(I)\,q}_{i'j',ij}(\omega) \equiv J^{\,q}_{i'j',ij}(\omega),
    \end{equation}
    and \eq{eq:EcMaestra_devJ} is approximated
    \begin{equation}\label{eq:EcMaestra_devJ_HT}
    \begin{split}
    \frac{d\Delta\sigma(\tau)}{d\tau} = -\frac{1}{2} \sum_q \sum_{i<j,i'<j'} J^{\,q}_{i'j',ij}(\omega_0) \bigg\{
    &\op{A}^{-q}_{i'j'}\,\op{A}^{q}_{ij}\,\Delta\sigma(\tau)
    - \op{A}^{q}_{ij}\,\Delta\sigma(\tau)\,\op{A}^{-q}_{i'j'}\\
    &+ \Delta\sigma(\tau)\,\op{A}^{q}_{ij}\,\op{A}^{-q}_{i'j'}\
    - \op{A}^{-q}_{i'j'}\,\Delta\sigma(\tau)\,\op{A}^{q}_{ij} \bigg\},
    \end{split}
    \end{equation}
    where we used $d\sigma_{eq}/d\tau = 0$; thus, $d\left(\Delta\sigma(\tau)\right)/d\tau = d\sigma(\tau)/d\tau$.\\

    Using the lineal equation \numeq{eq:EcMaestra_devJ_HT}, we can express the dynamics of the deviation density matrix elements in an orthonormal basis $\{\ket{m}\}$ (e.g. the preferred basis); this is
    \begin{equation}\label{eq:EcMaestra_devJ_HT_m}
    \begin{split}
    \frac{d\bra{m}\Delta\sigma(\tau)\ket{n}}{d\tau} = -\frac{1}{2} \sum_q \sum_{i<j,i'<j'} J^{\,q}_{i'j',ij}(\omega_0) \sum_{r,s}
    &\bra{m}\op{A}^{-q}_{i'j'}\ket{r}\bra{r}\op{A}^{q}_{ij}\ket{s}\\&\times\bra{s}\Delta\sigma(\tau)\ket{n} + \cdots.
    \end{split}
    \end{equation}

    In a spin system in contact with a thermal bath, the spin and environment operators are
    \begin{gather}
    \op{A}_{ij}^0 = \op{I}_{\op{z},i}\,\op{I}_{\op{z},j} - \frac{1}{4}\left(\op{I}_{+,i}\,\op{I}_{-,j} +
    \op{I}_{-,i}\,\op{I}_{+,j}\right),\quad
    \op{F}_{ij}^0 = \frac{\mu_0 \hbar \gamma_p^2}{4 \pi \; r_{ij}^3} (1- 3 \cos^2{\theta}_{ij}),\nonumber\\
    \op{A}_{ij}^{\pm1} = -\frac{3}{2}\left(\op{I}_{\op{z},i}\,\op{I}_{\pm,j} + \op{I}_{\pm,i}\,\op{I}_{\op{z},j}\right),\quad
    \op{F}_{ij}^{\pm1} = \frac{\mu_0 \hbar \gamma_p^2}{4 \pi \; r_{ij}^3} \sin{\theta}_{ij} \cos{\theta}_{ij} e^{\mp i \phi_{ij}},\label{eq:spin-latt_op}\\
    \op{A}_{ij}^{\pm2} = -\frac{3}{4}\op{I}_{\pm,i}\,\op{I}_{\pm,j},\quad
    \op{F}_{ij}^{\pm2}  = \frac{\mu_0 \hbar \gamma_p^2}{4 \pi \; r_{ij}^3} \sin^2{\theta}_{ij} e^{\mp 2i \phi_{ij}},\nonumber
    \end{gather}
    where $\theta_{ij}$ and $\phi_{ij}$ are the polar and azimuthal angles of the internuclear vector $\vec{r}_{ij}$ relative to a frame with the $\op{z}$-axis parallel to the external magnetic field, $r_{ij} = \abs{\vec{r}_{ij}}$ and $\gamma_p$ is the proton gyromagnetic ratio.
    The $\op{F}_{ij}^{\pm q}$ operators are represented by real functions in a semiclassical approximation.

    It is possible, in nematic LC, to show the following relationship \cite{SegBarGzzZar06}
    \begin{equation}\label{eq:rel_J}
    J^{\,q}_{i'j',ij}(\omega) =  \epsilon_{i'j'}\;\epsilon_{ij}\;J^{\,q}_{o,o}(\omega),
    \end{equation}
    with the index $o$ involving the spin ortho-pair in the indexes $i'j'$ and $ij$, and
    \begin{equation}\label{eq:def_ep}
    \epsilon_{ij} \equiv \overline{\op{F}^{\,0}_{ij}}/\overline{\op{F}^{\,0}_{o}}.
    \end{equation}

    Therefore, using \eq{eq:rel_J} in \eq{eq:EcMaestra_devJ_HT_m}, we obtain
    \begin{equation}\label{eq:EcMaestra_devJ_HT_m_Joo}
    \begin{split}
    \dot{\Delta\sigma}(\tau)_{m,n} = -\frac{1}{2} \sum_q J^{\,q}_{o,o}(\omega_0) \!\sum_{i<j,i'<j'}\! \epsilon_{i'j'}\,\epsilon_{ij} \sum_{r,s}\,
    &\bra{m}\op{A}^{-q}_{i'j'}\ket{r}\bra{r}\op{A}^{q}_{ij}\ket{s}\\&\times \Delta\sigma(\tau)_{s,n} + \cdots.
    \end{split}
    \end{equation}
    \eq{eq:EcMaestra_devJ_HT_m_Joo} is a linear equation with constant coefficients that relates a matrix element $\dot{\Delta\sigma}(\tau)_{m,n}$ with other elements $\Delta\sigma(\tau)_{m',n'}$. We can solve \numeq{eq:EcMaestra_devJ_HT_m_Joo} regarding $\dot{\Delta\sigma}(\tau)$ and $\Delta\sigma(\tau)$ as vectors, where the indexes $m,n$ are integrated in a unique index, and the coefficients form the elements of a linear transformation matrix between such vectors.
    Therefore, the solution of \eq{eq:EcMaestra_devJ_HT_m_Joo} can be obtained by diagonalization, where the eigenvalues of the transformation matrix are the decay rates of exponential functions.
    Such exponentials are the solutions for each vector element in the diagonalization eigenbasis; then, using this basis, we can obtain the relaxation dynamics for the elements of $\Delta\sigma(\tau)$ in the original basis. Finally, the evolution of the density matrix under relaxation is
    \begin{equation}\label{eq:sigma_rel_eq}
    \sigma(\tau) = \Delta\sigma(\tau) + \sigma_{eq}.
    \end{equation}

    We can numerically calculate \eq{eq:EcMaestra_devJ_HT_m_Joo} or \eq{eq:sigma_rel_eq} for a system of $N$ spins using an algorithm, but the computational strength growth exponentially with $N$. Nevertheless, we can use a representative few spins system in order to obtain the magnitude of the relaxation rates for a nematic LC molecule. Then, we perform such a calculation for four 1/2-spins located in the vertex of a rectangle with dimensions 2.45{\AA} $\times$ 4.23{\AA}, which represents the positions of the spins in a benzene ring. The dipolar couplings of this system are shown in \tabla{tab:4esp_DipCoup} and we use the values of the spectral densities for a PAA$_{d6}$ molecule \cite{SegBarGzzZar06} shown in \tabla{tab:Joo_PAAd6}.
    This calculation involves a density matrix of dimension $16 \times 16$, or its vectorial form with 256 elements, and a diagonalization of a symmetric coefficient matrix of dimension $256 \times 256$; thus, at most a set of 256 different eigenvalues or characteristic decay rates are obtained.\\

    \begin{table}[ht]
    	\centering
    	\begin{tabular}{c|c|c}
    		spin $i$ & spin $j$ & $\overline{\op{F}^{\,0}_{ij}}\,\unc{Hz}$\\
    		\hline
    		1 & 2 & -4083.6\\
    		\hline
    		3 & 4 & -4083.6\\
    		\hline
    		1 & 3 & 396.7\\
    		\hline
            2 & 4 & 396.7\\
    		\hline
            1 & 4 & 63.3\\
    		\hline
            2 & 3 & 63.3
    	\end{tabular}
    	\caption{Dipolar couplings for the four 1/2-spins model used in the numerical calculations. Ortho-pair coupling $\overline{\op{F}^{\,0}_{o}} = -4083.6\,\un{Hz}$ ($o \equiv 1,2 \equiv 3,4$).}
    	\label{tab:4esp_DipCoup}
    \end{table}

   \begin{table}[ht]
    	\centering
    	\begin{tabular}{c|c}
    		& $J^{\,q}_{o,o}(\omega_0)\,\unc{s^{-1}}$\\
    		\hline
    		$J^{\,0}_{o,o} =  J^{\,0\,{\rm ROT}}_{o,o}$ & 0.204\\
    		\hline
    		$J^{\,1}_{o,o} = J^{\,-1}_{o,o} = J^{\,1\,{\rm ROT}}_{o,o} + J^{\,1\,{\rm ODF}}_{o,o}$ & 0.034 + 0.267 = 0.301\\
    		\hline
    		$J^{\,2}_{o,o} = J^{\,-2}_{o,o} = J^{\,2\,{\rm ROT}}_{o,o}$ & 0.015
    	\end{tabular}
    	\caption{Spectral density values at $\nu_0 = 27\un{MHz}$ ($\nu_0 = \omega_0/(2\pi)$). ROT: Rotation, ODF: Order Director Fluctuation.}
    	\label{tab:Joo_PAAd6}
    \end{table}

    In order to show the effects of relaxation and decoherence over the signals, we calculate the evolution of the amplitude spectra for a single- and double-quantum multi-spin coherence under the $\mathcal{W}$-order initial condition of the density matrix. Such a condition is obtained immediately after the 2nd pulse of the Jeener-Broekaert sequence of \fig{fig:Puls} with $t_p = 81.4\un{\mu s}$ for the 4 1/2-spins model used. The dynamics for a time $\tau$ after the 2nd pulse is calculated in a free-evolution scheme (without the reversion block) affected only by relaxation, using the result of \eq{eq:EcMaestra_devJ_HT_m_Joo}, or by decoherence, using the eigen-selective decoherence function of \eq{eq:G_mn_LC_aproxfree} with a Gaussian form \bib{SegZam11,SegZam13,SegGzzZam_Annals21}.
    In \figs{fig:Evol_Esp_RD_C1} and \numfig{fig:Evol_Esp_RD_C2}, we show such a dynamics for the coherence amplitude spectrum obtained by the projection of the operator $\op{I}_{\op{y}} + \op{T}_{2,+1} + \op{T}_{2,-1}$ (single-quantum multi-spin) and $\op{T}_{2,+2} + \op{T}_{2,-2}$ (double-quantum multi-spin) on the density matrix evolution, where $\op{I}_{\op{y}} \equiv \sum_{i=1}^{4} \op{I}_{\op{y},i}$, $\op{T}_{2,\pm1} \equiv \sum_{i=1}^{4}\sum_{j=i+1}^{4} \op{A}_{ij}^{\pm1}$, and $\op{T}_{2,\pm2} = \sum_{i=1}^{4}\sum_{j=i+1}^{4} \op{A}_{ij}^{\pm2}$.\\

    \insertfig{ht}{fig:Evol_Esp_RD_C1}{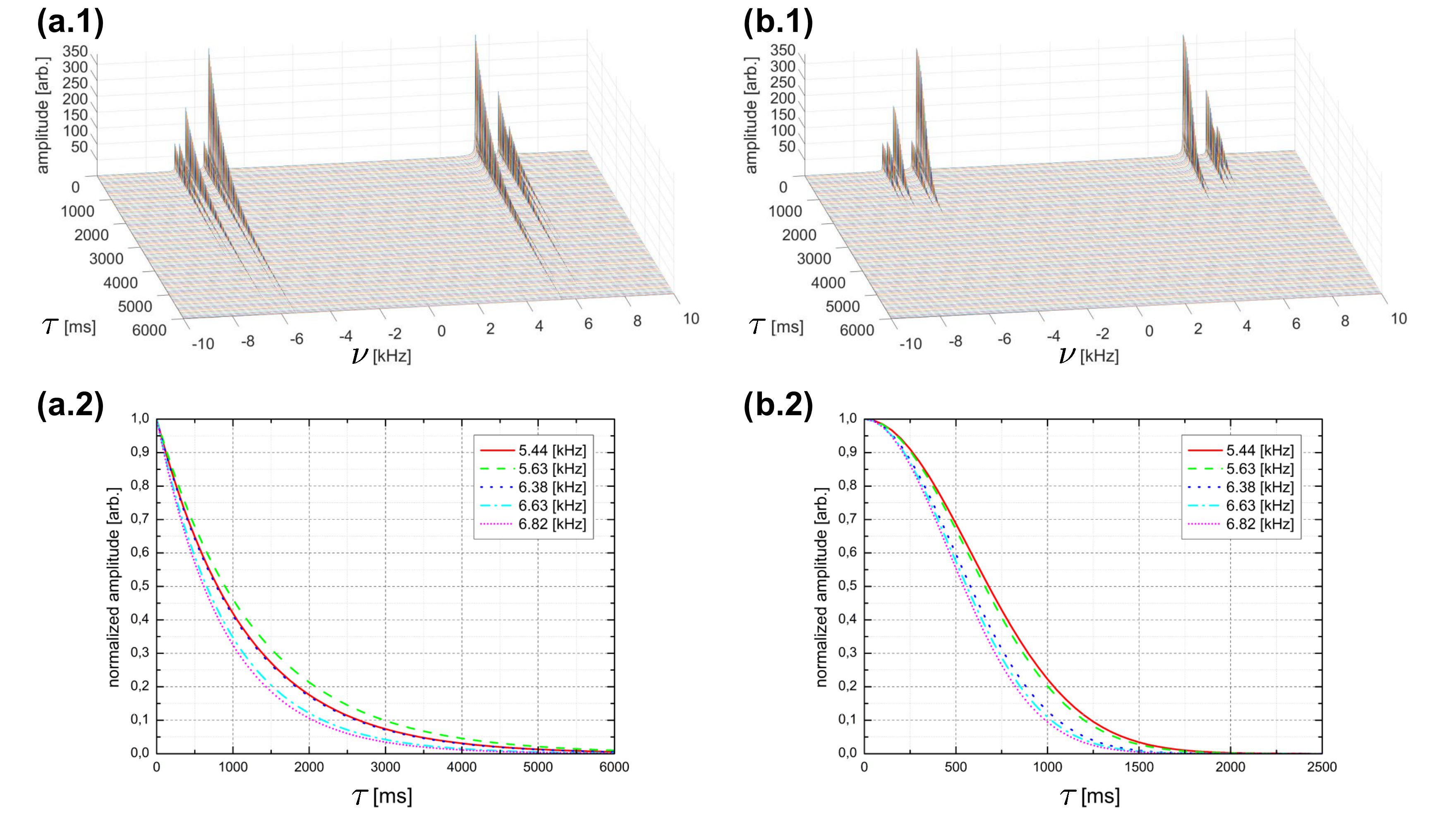}{15.0}{8.56}{Comparison of the amplitude of single-quantum multi-spin spectrum produced by $\op{I}_{\op{y}} + \op{T}_{2,+1} + \op{T}_{2,-1}$ under relaxation (a.1) and decoherence (b.1) for a 4 1/2-spins system in the $\mathcal{W}$-order initial condition, $t_p = 81.4\un{\mu s}$ and evolution in the time $\tau$. Evolution of the normalized amplitude of spectral lines; (a.2) relaxation and (b.2) decoherence.}

    \insertfig{ht}{fig:Evol_Esp_RD_C2}{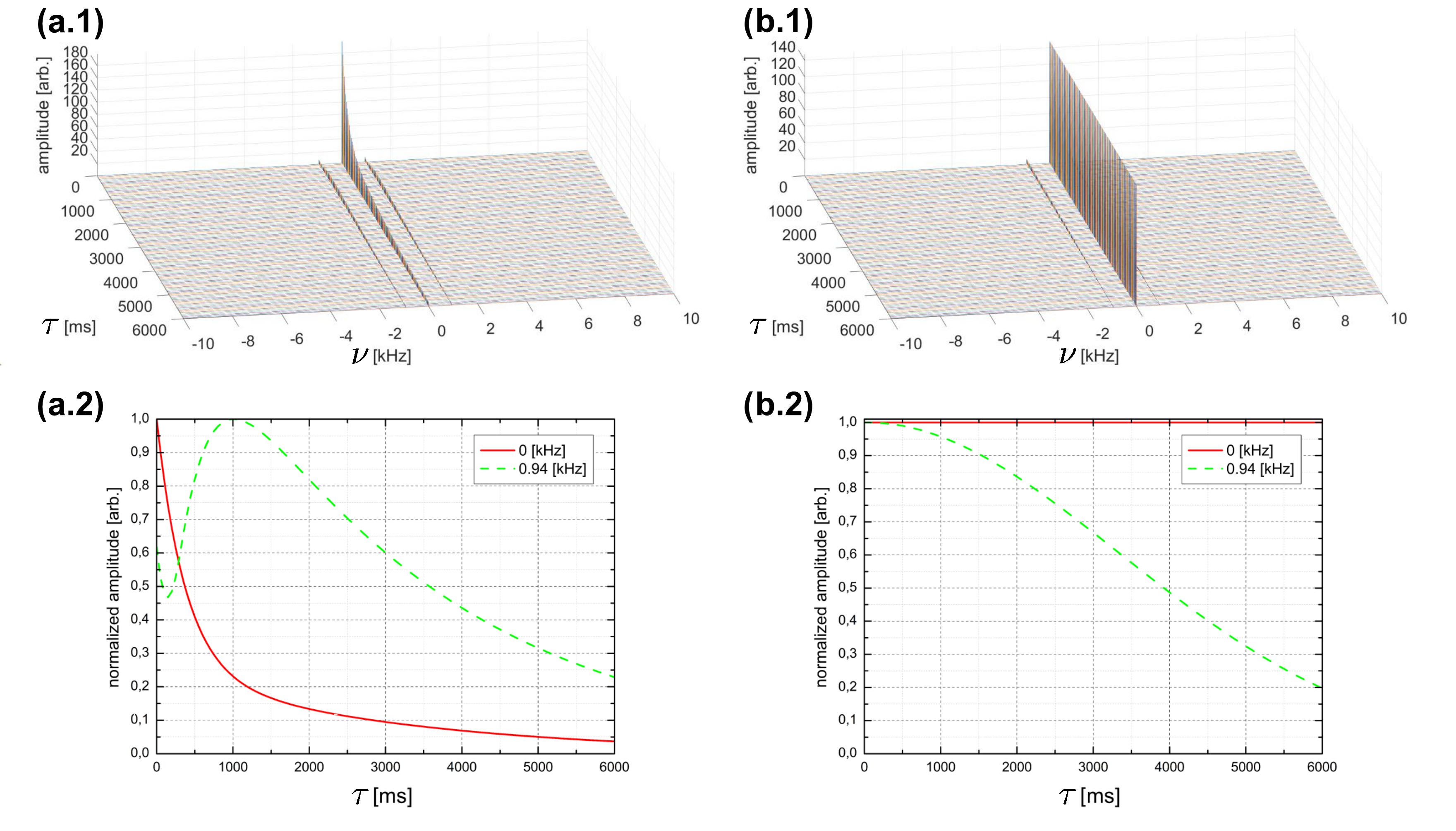}{15.0}{8.56}{Comparison of the amplitude of double-quantum multi-spin spectrum produced by $\op{T}_{2,+2} + \op{T}_{2,-2}$ under relaxation (a.1) and decoherence (b.1) for a 4 1/2-spins system in the $\mathcal{W}$-order initial condition, $t_p = 81.4\un{\mu s}$ and evolution in the time $\tau$. Evolution of the normalized amplitude of spectral lines; (a.2) relaxation and (b.2) decoherence.}

    We can see in (a.2) and (b.2) of \figs{fig:Evol_Esp_RD_C1} and \numfig{fig:Evol_Esp_RD_C2} the evolution of different frequency lines of the amplitude spectra shown respectively in (a.1) and (b.1), where such lines are normalized taking their maximum values to easily compare the decay time between them.
    The spectral lines in (a.2) of \figs{fig:Evol_Esp_RD_C1} and \numfig{fig:Evol_Esp_RD_C2} show that relaxation is not eigen-selective (see the text below \eq{eq:G_mn_LC}), besides the single-quantum coherence lines present characteristic decay times around $1\un{s}$ (see \fig{fig:Evol_Esp_RD_C1} (a.2)). In particular, in \fig{fig:Evol_Esp_RD_C1} (a.2) the frequency line at $6.38\un{kHz}$ has practically the equal decay time than at $5.44\un{kHz}$, and the line at $5.63\un{kHz}$ decay slowly than at $5.44\un{kHz}$ (in a inverse relationship that under the effect of eigen-selectivity).
    Moreover, in \fig{fig:Evol_Esp_RD_C2} (a.2) the zero frequency line has a similar decay time than the lines in \fig{fig:Evol_Esp_RD_C1} (a.2), which is completely different that in the eigen-selective case (where the zero line does not decay), and the line at $0.94\un{kHz}$ presents a non-monotonic decay.

    The decoherence process was studied in nematic LC by mean of single-quantum coherence or FID experiments \bib{SegZam11,SegZam13} and a theory was proposed to explain the measured decay times of around $1\un{ms}$ of the observables, but a deduction of this time scale from the physical parameters of the environment is not done yet. Nevertheless, in a recent work \bib{SegGzzZam_Annals21} such a time scale was deduced from the environment parameters in a system of non-interacting spin pairs coupled with a common phonon bath, where the obtained fast-decoherence function is temperature independent. Therefore, it is plausible that in nematic LC the decoherence process has a similar nature that in such a work with a characteristic decay time in the order of $1\un{ms}$. Then, with the aim of qualitatively comparing the decoherence and relaxation processes, we use in the calculation of the matrix density evolution the following Gaussian decoherence function
    \begin{equation}\label{eq:G_mn_calcDeco}
        G_{mn}(\tau) = e^{-\left[2 \left(E_m - E_n\right)/\abs{E}_M\right]^2(\tau/\tau_D)^2},
    \end{equation}
    where $E_m$ is the eigenvalue of the secular dipolar Hamiltonian associated with the eigenstate of the $m$th row o column of the density matrix, $\abs{E}_M$ is the maximum absolute value of the eingenvalues $E_m$, and $\tau_D$ is the characteristic decay time. Besides, we use the average decay time of the spectral lines of \fig{fig:Evol_Esp_RD_C1} (a.2) to assign the value of the decoherence decay time, this is $\tau_D = 1083\un{ms}$.\\
    Applying \eq{eq:G_mn_calcDeco} on the initial condition for the density matrix we obtain our dynamics under decoherence to compare with relaxation.
    The evolution of the single- and double-quantum coherence spectra under decoherence is shown in (b.1) of \fig{fig:Evol_Esp_RD_C1} and \fig{fig:Evol_Esp_RD_C2}, respectively. We observe in \fig{fig:Evol_Esp_RD_C1} (b.2) the expected eigen-selective behavior, where the higher the frequency of the spectral line, the shorter the decay time. Moreover, in \fig{fig:Evol_Esp_RD_C2} (b.2) we can see that the zero frequency line is not affected by the adiabatic quantum decoherence. This last property is necessary for the build-up of quasi-equilibrium states; we discuss about that in the following.\\

    A direct comparison of the matrix elements of the density operator under both processes may be helpful to understand the effect of eigen-selectivity over the features of the state evolution and to clarify the difference between them. In \fig{fig:Evol_MatD_RD} we show the evolution of the amplitude of the matrix elements under relaxation and decoherence for the 4 1/2-spins system, starting from the $\mathcal{W}$-order initial condition. The matrixes have dimension $16 \times 16$ and they correspond to the deviation matrix
    \begin{equation}\label{eq:OpMatDens_RD}
        \widetilde{\Delta}\sigma(\tau) \equiv \left(\sigma(\tau) - \frac{\op{1}}{2^4}\right)\frac{2^4}{\beta_0\,\omega_0},
    \end{equation}
    where $\omega_0 \equiv 2\pi\,\nu_0$ is the Larmor frequency (with $\nu_0$ in unit of $\un{Hz}$). The operator \numeq{eq:OpMatDens_RD} is the part of the density matrix which affects to the observables in NMR experiments and the absolute value of its elements ranges $0 \leq \abs{\widetilde{\Delta}\sigma_{m,n}(\tau)} \leq 2$ for 4 1/2-spins (the value 2 correspond to the maximum absolute value of the eigenvalues of $\op{I}_{\op{z}}$). Besides, we write the matrixes using the eigenbasis of the secular dipolar Hamiltonian, ordering the row and column indexes with their eigenvalues; thus, in the main diagonal there are blocks or submatrixes which correspond to degenerate spaces.\\

    \insertfig{ht}{fig:Evol_MatD_RD}{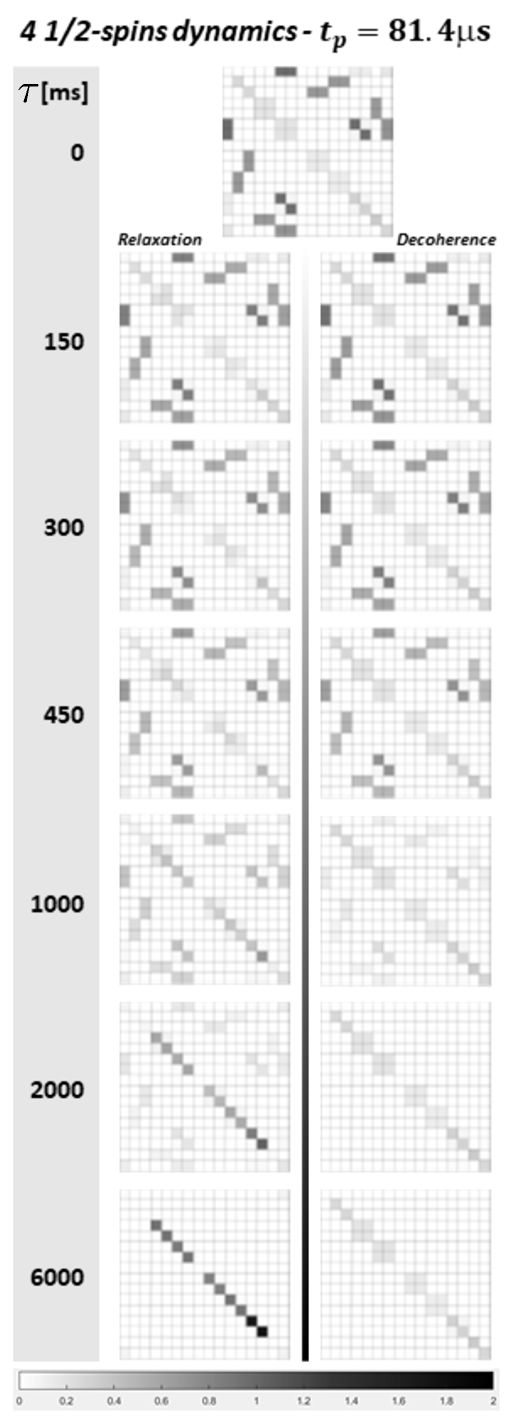}{4.32}{12.0}{Comparison of the matrix density amplitude dynamics of the deviation matrix $\widetilde{\Delta}\sigma(\tau)$ under relaxation and decoherence for a 4 1/2-spins system in the $\mathcal{W}$-order initial condition, $t_p = 81.4\un{\mu s}$ and evolution in the time $\tau$.}

    We can observe in both processes that the elements out of the main diagonal vanish for a long time $\tau$. Nevertheless, in the dynamics under relaxation, the final form is strictly diagonal and the degenerate blocks are not preserved. Moreover, such a final form corresponds to the operator $\widetilde{\Delta}\sigma(\tau) = \op{I}_{\op{z}}$ (or $\sigma(\tau) = \sigma_{eq}$) and the density operator attains the thermal equilibrium. On the other hand, the decoherence process preserves the elements of the submatrixes in the space of the density operator belonging to the dipolar energy degenerate eigenvalues. Such a diagonal-in-block space can be developed as a sum of free-decoherence operators which are the quasiinvariants of the dynamics, then the dynamics under decoherence attains a quasiequilibrium state.\\

    Finally, we can conclude that a relaxation process, which is governed by a master equation theory, can not attain an intermediate or quasiequilibrium state before the thermal equilibrium condition. Besides, the relaxation time scale measured and calculated in nematic LC is around 3 orders of magnitude bigger than the observed in the decay time scale of coherence. Otherwise, a first principle theory \bib{SegGzzZam_Annals21} for an adiabatic quantum decoherence process predicts a fast decay time (with the same magnitude order as in NMR experiments) for coherences or the out-of-diagonal elements of the density matrix, with a characteristic eigen-selective effect that preserves a diagonal-in-block form of the density operator. Then, the decoherence process is quantitatively and qualitatively compatible with the build-up time scale of quasiinvariant states observed in nematic LC and solids in NMR experiments. Moreover, we can infer that non-conservative energy relaxation processes drive a quasiequilibrium state toward a final thermal equilibrium density matrix in a long time scale; where the quasiinvariants are builded-up by conservative energy adiabatic decoherence processes in an earlier time scale.\\
    
    In the next section, we show the results of the proposed experiment of \fig{fig:Puls} applied to different nematic LC samples.

\section{Experimental setup and results}\label{se:QED_ExpSetMeas}

    In this section we present the experimental results on LC samples in the nematic phase, obtained with the pulse sequences shown in \fig{fig:Puls}, and described in \sect{se:QEDetect}.
    The experiments which use the reversion sequence MREV8 were carried out on samples of 5CB (4'-pentyl-4-biphenyl-carbonitrile) and PAAd$_6$ (methyl deuterated p-azoxyanisole), using a home-built pulsed spectrometer, based on a Varian EM360 magnet ($60\un{MHz}$ for protons). The electronic setup allows control of the pulse phases, homogeneity of the magnetic field, and sample temperature with stability of $\pm0.1\un{^oC}$. Experiments which use the reversion sequence MS were conducted on a Bruker minispec mq20 ($20\un{MHz}$) on 8CB (4'-octyl-4-biphenyl-carbonitrile).

    The time $t_{p\mathcal{S}}$ needed to prepare the so-called `strong' $\mathcal{S}$-order corresponds to the maximum slope of the FID signal (FID after a ($\pi/2$) pulse). The $\mathcal{S}$ signal so prepared with the JB sequence has the shape of the FID time derivative \cite{bulju09}. Typically in nematic LC, it oscillates with a frequency related to the stronger dipolar couplings within the molecule, and its amplitude is about 25\% that of the FID.  The time $t_{p\mathcal{W}}$ to prepare `weak' or $\mathcal{W}$-order \cite{MenGzzZam05,Bonin13} corresponds to the first node of the $\mathcal{S}$ signal. The $\mathcal{W}$ signal amplitude is still smaller, about 10\% the FID.
    Table \ref{tab:QE_time} shows the preparation times used in the three samples.

    \begin{table}[ht]
    	\centering
    	\begin{tabular}{c|c|c}
    		& $t_{p\mathcal{S}}$ & $t_{p\mathcal{W}}$ \\
    		\hline
    		5CB & $27\un{\mu s}$ & $69\un{\mu s}$ \\
    		\hline
    		8CB & $34\un{\mu s}$ & $72\un{\mu s}$ \\
    		\hline
    		PAAd$_6$ & $40\un{\mu s}$ & $84\un{\mu s}$ \\
    	\end{tabular}
    	\caption{Time between the JB pulse pair to prepare the $\mathcal{S}$ and $\mathcal{W}$ states.}
    	\label{tab:QE_time}
    \end{table}

    Figure \ref{fig:ExpCM_5CB_S} shows the experimental results on 5CB with $t_p = t_{p\mathcal{S}}$. The data are organized so that the $\mu$-axis displays the spectra of the different coherence orders  $\eta$ ($\eta = 0, \pm 1, \pm2, \ldots $). The spectrum of \mbox{$\eta$-order} is symmetric to the one with \mbox{$-\eta$-order}, and the figure is arranged to show all the resolved coherences.
    We chose a step time ($\Delta t$) of $t$ that shows all the spectral frequencies of each coherence ($\nu_M = 1/(2\,\Delta t)$) and a maximum time value ($t_M$) large enough to assure that the coherences vanished by decoherence.
    It is important to note a trade-off condition between $\Delta t$ and $t_M$ because we need more experimental effort (run more sequences) to shorten the former or lengthen the latter.
    In order to optimize the reversion strategies, we first compare the performance of a single MREV8 block with variable delays and that of a train of MREV8 blocks. The $\tau$-axis shows the spectra measured.
    Figure \ref{fig:ExpCM_5CB_S}(a) corresponds to the single-MREV8 sequence with $\tau$ increasing in steps of $120\un{\mu s}$ and (b) to the block train with $\tau = n\times 90.36\un{\mu s}$. In both cases, the largest contribution is that of the zero-order coherence, and, as expected, its spectrum is sinc-shaped (because of inevitable truncation of a zero frequency signal). A much lower amplitude single-quantum spectrum, observable at short times, is seen to attenuate for $\tau \sim 400\un{\mu s}$.
    As seen in \fig{fig:ExpCM_5CB_S} both sequences encode the same coherences; however, the block train yields better results because the zero-order coherence amplitude remains visible at long $\tau$. On the contrary, the other sequence spoils the signal within the same period. Thus in the subsequent experiments on PAA$d_6$, we use the series of MREV8 blocks.

    In \apen{ap:non-ideal}, we analyze the effect of non-idealities in the MREV8 sequence. We calculate the contribution of the dipolar Hamiltonian's non-secular terms, which emerges for long evolution times under $\opc H_S$.
    The $D$-block performance depends both on  the initial state which it is applied to and on the time $\tau_1$ (see \fig{fig:Puls}(b)): shorter $\tau_1$ imply better performance.
    Since the MREV8 train involves many pulses, the dipolar evolution during rf irradiation also becomes relevant. To mitigate the pulse width effect, we optimized the sequence by compensating the $\pi/2$ pulse length and changed the relation 1:2 of the delays shown in \fig{fig:Puls}(b)\cite{SegZam13}.
    The shortest optimized experimental times $\tau_c$ obtained are  $90.36\un{\mu s}$ for 5CB and $88.36\un{\mu s}$ for PAA$d_6$; thus, the total reversion time is $\tau = n\,\tau_c$, with $n$ the number of blocks.
    The single MREV8 block experiments use an optimized block and the short-to-long time between pulses quotient varies proportionally as $\tau$ increases.
    It is important to note that any misadjustments in the reversion sequence can affect the spectra amplitude evolution; however, they would never mimic the effect of eigen-selection (see Section II.D of Ref.\cite{SegZam13} and supplementary material).

    \insertfig{ht}{fig:ExpCM_5CB_S}{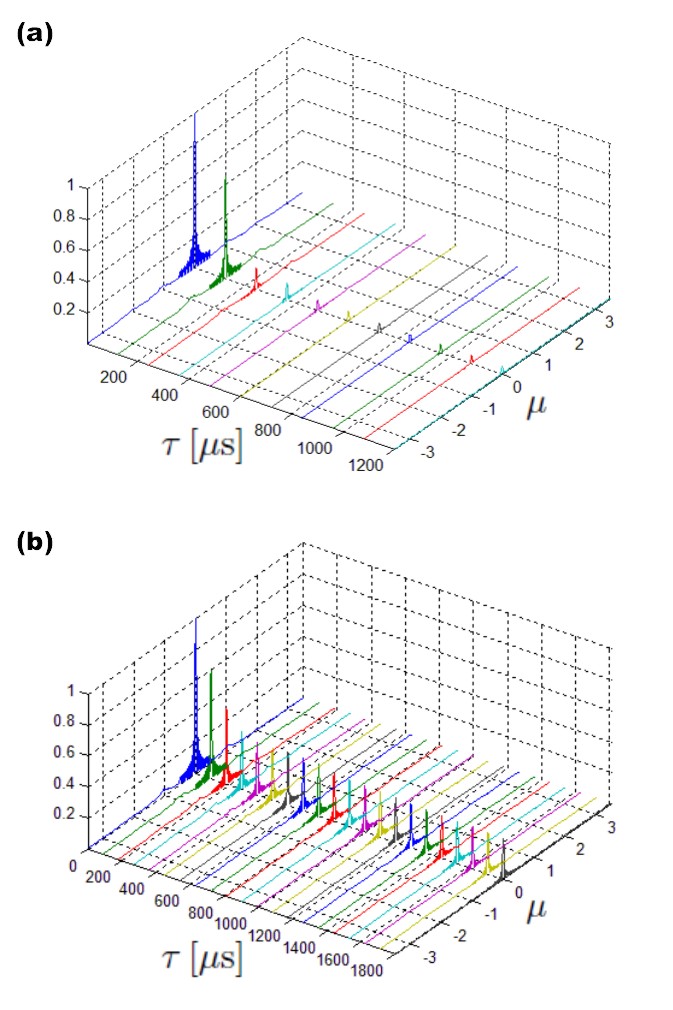}{9.0}{13.50}{Experimental spectra of the coherences measured in nematic 5CB. The time $t_p$ is set to prepare the strong or $\mathcal{S}$-order ($t_p = 27\un{\mu s}$). The $\mu$-axis shows the series of MQC spectra (symmetric with respect to zero-order).     Each spectrum is limited to $\nu_M = 25\un{kHz}$ and to a maximum coherence time $t_M = 400\un{\mu s}$.     The reversion block in (a) is a single MREV8 sequence with total reversion time $\tau$ varied in steps of $120\un{\mu s}$; (b) increasing number of short blocks with a characteristic step-time of $\tau_c = 90.36\un{\mu s}$.}

    A richer coherence spectrum arises by using $t_{p\mathcal{W}}$ to prepare the initial state. \fig{fig:ExpCM_5CB_W} shows that the \mbox{zero-}, \mbox{single-} and double-quantum coherence spectra are well distinguished.
    This agrees with previous experiments that use encoding MQC on two orthogonal bases to monitor the proton spin dynamics after the JB pulse pair in nematic 5CB \cite{bulju09}. That experiment showed that after $t \sim 300\un{\mu s}$ the tensor structure of the $\mathcal{S}$-state corresponds to two-spin dipolar order. Contrastingly, the $\mathcal{W}$-state produces discernible signals of multi-spin order, indicating that its tensor structure must involve products of individual angular momentum operators of many spins.
    The zero-quantum coherence amplitude in \fig{fig:ExpCM_5CB_W} shows a slow undulation with $\tau$ similar to that shown in \fig{fig:CalcCM_PAAd6}(c2).
    This fact suggests a non-ideal performance of the reversion sequence (see \apen{ap:non-ideal}).
    We should also notice that the single-quantum coherence attenuates much faster than the simulation and the double-quantum coherence goes even faster (instead of growing slowly as in \fig{fig:CalcCM_PAAd6}(c2)). Thus the monotonous attenuation of single and double-quantum coherences cannot be ascribed to an experimental artifact.

    \insertfig{ht}{fig:ExpCM_5CB_W}{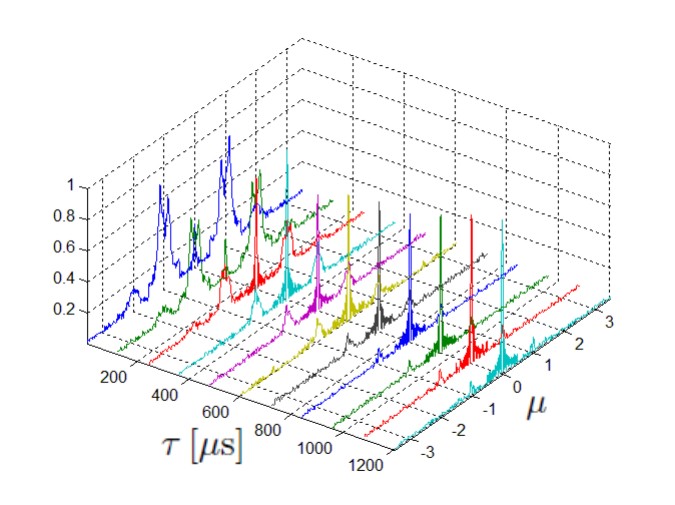}{9.0}{6.75}{Experimental spectra of the coherences measured in nematic 5CB. The time $t_p$ is set to prepare the weak or $\mathcal{W}$-order ($t_p = 69\un{\mu s}$). The $\mu$-axis shows the series of MQC spectra (symmetric with respect to zero-order).
    Each spectrum is limited to $\nu_M = 25\un{kHz}$ and to a maximum coherence time $t_M = 400\un{\mu s}$.
    The reversion block corresponds to a single MREV8 sequence where the total reversion time $\tau$ is varied in steps of $120\un{\mu s}$.}

    The evolution of the multi-spin single-quantum coherence deserves a separate analysis: \fig{fig:C1_5CB_W}(a) shows a stack plot of the single-quantum spectra from \fig{fig:ExpCM_5CB_W} ($\mathcal{W}$ condition) at different values of $\tau$.
    Notice that, the shape of these spectra correspond to \eq{eq:nq_cspec} with $\zeta > 1$ and $\mu=1$.
    It is then expectable that they differ from the FID spectrum, which corresponds to the single-spin, single-quantum coherence with $\zeta=\Lambda=\mu=1$.
    To display the frequency content as a function of the reversion time, \fig{fig:C1_5CB_W}(b) shows the amplitudes at some fixed frequency values, normalized to the corresponding maximum value.
    Following Refs.\cite{SegZam11,SegZam13} we fit the amplitude decay data with Gaussian functions (dashed curves)
    \begin{equation}\label{eq:gauss_fit}
        y(\tau) = y_0 + A\,e^{-(\tau/\tau_d)^2},
    \end{equation}
    with the fitting variables $y_0$, $A$ and $\tau_d$.
    The occurrence of a zero-frequency component may indicate the presence of one, or a combination, of tensor operators that belong to the commutative space of $\H_{SE}$, and thus are immune to adiabatic decoherence. The fact that we observe they attenuate with $\tau$ can obey to the convolution with the decaying low-frequency components.
    The parameter $y_0$ in \numeq{eq:gauss_fit} is used to better discern the frequency selective attenuation.
    The characteristic decay times $\tau_d$ (inset of the same figure) are noticeably shorter for higher frequencies. The highest frequency decay time is approximately one half the one at the lowest measured frequency.
    Previous work on 5CB \cite{SegZam13} and 8CB \cite{GzzSegZam11} gave a first indication that the attenuation of the overall amplitude of the reverted single-spin, single-quantum signals is a consequence of irreversible decoherence.
    The result shown in \fig{fig:C1_5CB_W} confirms this fact and also indicates that the frequency components are selectively attenuated during the build-up of quasi-equilibrium.
    The double-quantum peaks have a much smaller amplitude than the multi-spin single-quantum ones, just as happens with the calculated spectra (see \apen{ap:non-ideal}), and their S/N ratio is too small to analyze the selective narrowing. However, it can be seen that their amplitude decays monotonously with $\tau$.

    \insertfig{ht}{fig:C1_5CB_W}{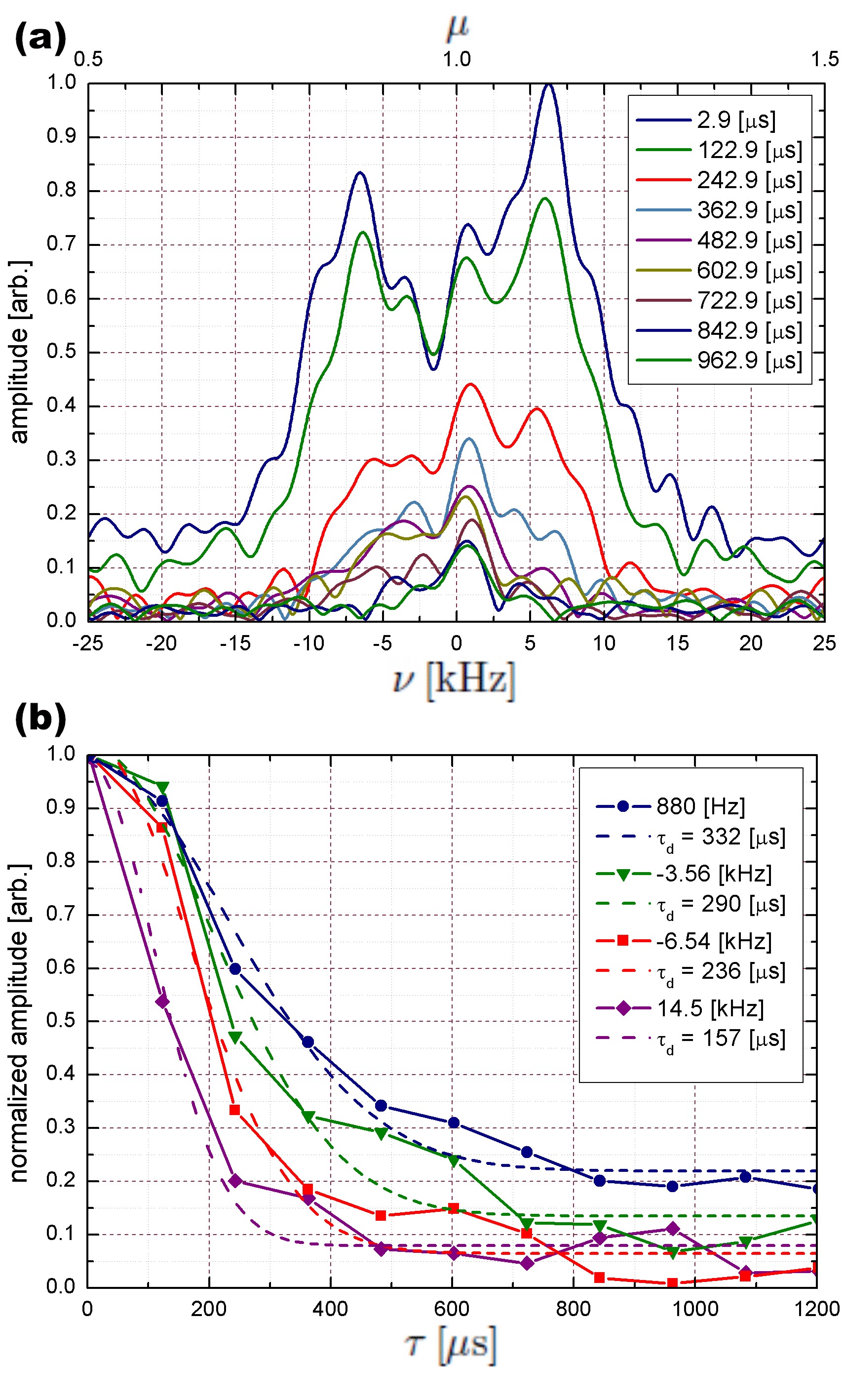}{8.0}{13.3}{(a) Front view of the multi-spin, single-quantum coherence ($\mu=1$) spectra of \fig{fig:ExpCM_5CB_W}, for the $\cal W$-order on nematic 5CB. (b) Time evolution of the (normalized) amplitude at different frequencies on the spectra of (a). Dashed lines are Gaussian fittings with the characteristic times $\tau_d$ shown in the inset box.}

    The behavior of PAA$_{d6}$ under the same reversion sequence (concatenated blocks) is shown in \fig{fig:ExpCM_PAAd6}, for the nematic phase at $T = 115\un{^oC}$, where (a) corresponds to a preparation time for the $\mathcal{S}$ condition and (b) to the $\mathcal{W}$ condition; again, the $\mathcal{W}$ condition gives rise to a richest coherence content, associated with the multi-spin tensor structure of the state.
    The single-quantum spectrum in \fig{fig:ExpCM_PAAd6}(b) is a well resolved doublet with no component centered at zero frequency.
    The overall attenuation time of the spectra with $\tau$ is in the same range as 5CB but is considerably larger.
    In this compound we cannot distinguish the frequency selective narrowing found in 5CB (and also 8CB, see \fig{fig:C1_8CB}) and both wings decay with the same rate. Within the AQD approach we link this fact with the absence of low or zero frequency components that, being less susceptible to irreversible decoherence would serve as a gauge for comparison with the higher frequency components.
    The zero-order coherence spectrum in \fig{fig:ExpCM_PAAd6}(b) is long lived and the $\tau$ dependence of its amplitude in both plots (a) and (b) can be explained by the inherent non-ideality of the reversion sequence (see \fig{fig:CalcCM_PAAd6}(c2) in \apen{ap:non-ideal}).
    Therefore, the eigen-selectivity can be shown by comparing the lived time under reversion of the single- and zero-order coherence spectrum.

    \insertfig{ht}{fig:ExpCM_PAAd6}{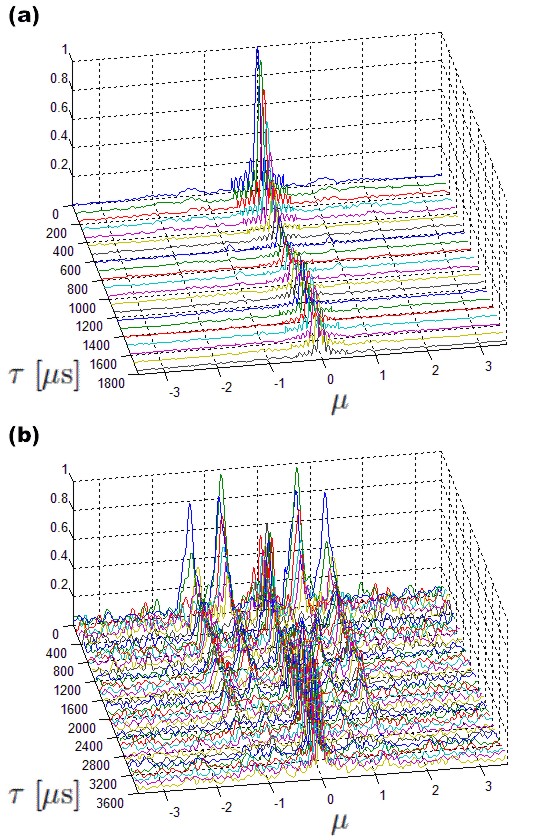}{9.0}{13.50}{Experimental spectra of the coherences measured in nematic PAA$_{d6}$ at $T = 115\un{^oC}$, using blocks of MREV8 sequences of duration $\tau_c = 88.36\un{\mu s}$, which define the $\tau$ axis step-times. The time $t_p$ is fixed to set different dipolar QE states, (a) Strong or $\mathcal{S}$-order ($t_p = 40\un{\mu s}$), (b) Weaker or $\mathcal{W}$-order ($t_p = 84\un{\mu s}$). The $\mu$-axis shows the series of MQC spectra (symmetric with respect to zero-order).     In (a) each coherence spectrum is limited to $\nu_M = 5\un{kHz}$ and to a maximum coherence time $t_M = 1000\un{\mu s}$;    in (b) to $\nu_M = 10\un{kHz}$ and $t_M = 500\un{\mu s}$.}

    In order to show that the frequency-selective amplitude decay is independent of the particular reversion sequence and compound, we carried out a similar experiment using the reversion sequence known as Magic Sandwich (MS), on 8CB in the nematic phase.
    In this case  $\tau = 1.5\,\tau_M$ as in \fig{fig:Puls}(c) \cite{GzzSegZam11}.
    Figure \ref{fig:C1_8CB}(a) shows the multi-spin single-quantum spectra at different $D$ block lengths, with the initial state prepared in the $\mathcal{W}$ condition, and (b) shows that the higher frequency component amplitudes attenuate with $\tau$ notably faster than the central peak, similarly to 5CB. By fitting these curves with Gaussian functions as \eq{eq:gauss_fit}, we get the characteristic decay times: $2\un{ms}$ and $1.2\un{ms}$.

    \insertfig{ht}{fig:C1_8CB}{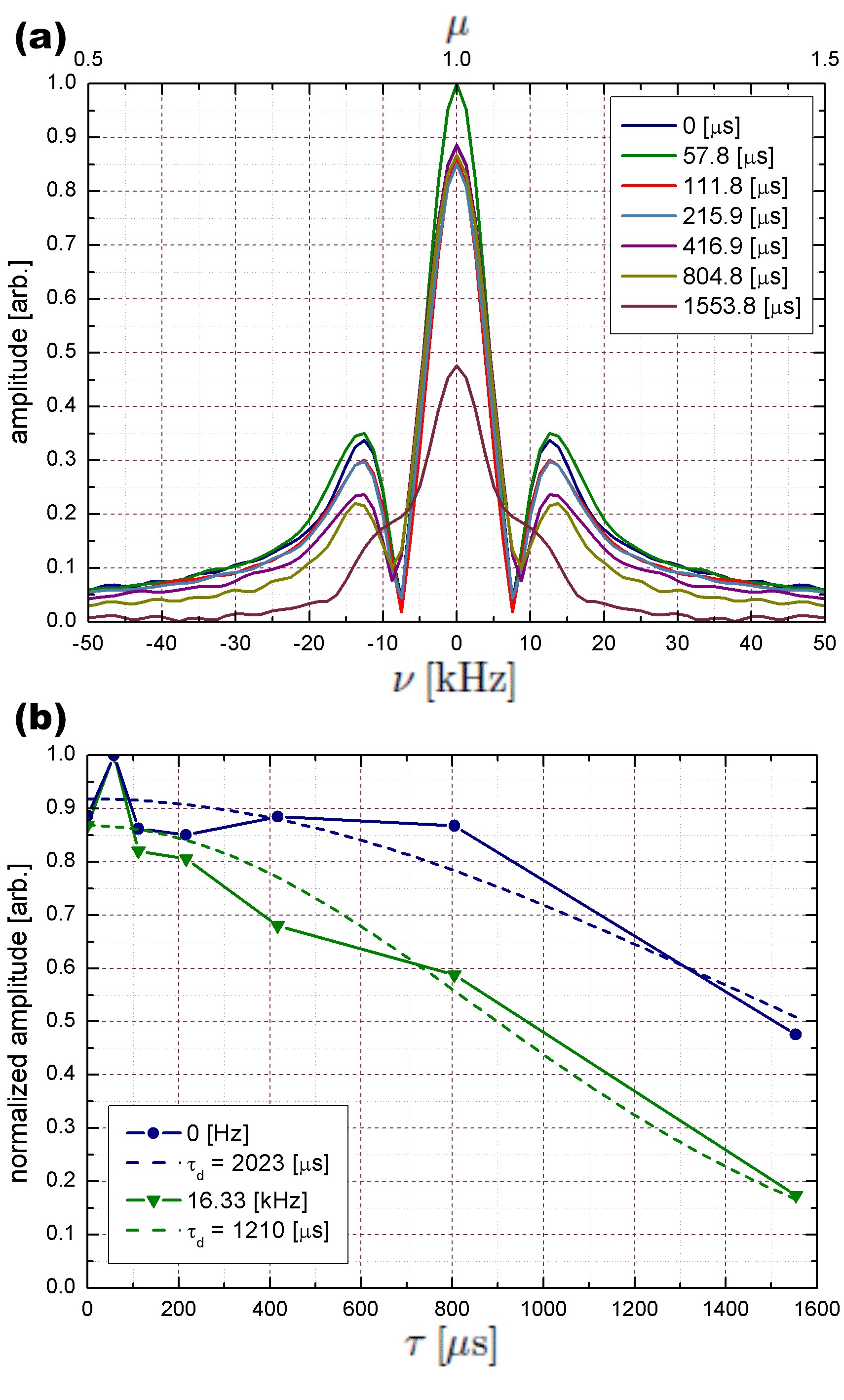}{8.0}{13.3}{(a) Stack plot of the spectra of single-quantum coherence for $\mathcal{W}$-order ($t_p = 72\un{\mu s}$) in 8CB for different reversion times $\tau$. (b) Time evolution of the (normalized) amplitude of the central  peak (blue circles) and the side peaks (green triangles). The dashed lines are Gaussian fittings with the characteristic times $\tau_d$ as are shown in the inset box.}

\section{Discussion and conclusions}\label{se:Conclu}

    A variety of nuclear spin systems develop QE states in an early timescale; that is, in short times as compared with the lapse necessary to transfer a significant amount of energy to the environment. A practical NMR experiment useful to prepare such states is the Jeener-Broekaert pulse sequence. We prepare initial coherent states in nematic liquid crystals with the JB pulse pair and analyze their attenuation in this work.
    We concentrate on the multi-spin coherence spectra inherent to the so-called $\cal W$ initial condition \cite{SegBonGzzAcoZam09,Bonin13}. The experiment allowed monitoring their evolution while refocusing the dipolar dynamics. The \mbox{single-} and double-quantum coherence amplitudes attenuate monotonously under different reversion schemes, in a way that definitely differs from the evolution of a closed system under non-ideal experimental conditions. Besides, the frequency content of the multi-spin single-quantum coherence of the three studied compounds changes with the reversion time, showing that the higher-frequency components attenuate faster than the lower-frequency ones. We interpret these results as new evidence in favor of eigen-selective decoherence as the mechanism that drives the spin system from an initial state to QE.

    In solid-state NMR literature, the spin system is frequently considered as strictly closed in the early timescale, and spin-lattice relaxation is viewed as the only channel by which the environment acts on the spin system \cite{multispBoutis12,Emsley12}. In this vision, QE is not an actual state but merely an apparent effect on the expectation values produced by the interference of phase factors. Following this vision, the decay of spin observables under reversion experiments is attributed either to imperfections in the pulse sequence settings or dissipative interaction among all the spins in the sample. Ideally, after improving the experimental techniques, the only limit to reversion would be `internal equilibration' or `pseudo-thermalization', a unitary process that would cause the observed coherence decay.
    Although this line of thought might, in principle, give a phenomenologic explanation of the observed signal attenuation in reversion experiments in solids, it cannot explain the phenomenon in nematic LC. In the LC mesophases only the few spins at each molecule are effectively interacting through their magnetic degrees of freedom. This fact rules out the argument of `interaction among a thermodynamic number of spins' as the mechanism inhibiting longer experimental reversion times.

    On the other hand, in the field of open quantum systems, adiabatic decoherence due to the quantum system-environment coupling is shown to be responsible for the decay of the matrix elements of the density operator \cite{SegZam11,SegZam13,SegGzzZam_Annals21,DomGzzSegZam16,Yukalov11,privman98,DomZamSegGzz17}.
    This approach does not rely on the hypothesis of thermodynamic-sized systems.
    Particularly, in Refs.\cite{SegZam11,SegZam13} the attenuation of single-spin NMR signals in reversion experiments on nematic LC is explained under the formalism of open quantum systems. Such attenuation is essentially irreversible, even in the assumption of a perfect-reversion experiment.
    In this way, the irreversible decoherence that leads evolution within the intermediate time scale, with a selective frequency attenuation or eigen-selectivity effect, arises as a mechanism to explain the development of QE. Since the final QE state depends on the initial condition $\rho_S(0)$ (see \eq{eq:sigma_isol}), AQD naturally explains the experimentally observed occurrence of different QE states in LC.

    Thus, the evidence shown in \sect{se:QED_ExpSetMeas} that multi-spin coherences in LC samples also carry the signature of adiabatic decoherence leads us to conclude that QE states are actual states represented by a diagonal-in-blocks density matrix on a preferred basis.
    Accordingly, they admit an expansion in multiple constants of motion (quasi-invariants), which implies a convenient analytical representation from a theoretical viewpoint.
    Since such quasi-invariants are immune to decoherence, they may be posed as candidates to memory units in quantum computations; also, since they only evolve due to relaxation, they are useful as sensors of the molecular motions and other long-timescale effects of the environment over the spin dynamics.

\section{Acknowledgement}

    This work was supported by Secretar\'ia de Ciencia y T\'ecnica, Universidad Nacional de C\'ordoba and MINCyT C\'ordoba. H.H.S. thanks CONICET for financial support.



\begin{thebibliography}{10}
\bibitem{footnote_T2}
    In principle, the spectral densities $J_0$, $J_1$ and $J_2$ contribute to $T_2$
      in the isotropic phase and might change their relative weights in the nematic
      phase, however, the mentioned experiment \cite{GzzSegZam11} was carried on at
      $20\un{MHz}$ where the order fluctuations of the director (OFD) contribute
      negligibly. This fact allows assuming that if it were possible to measure
      $T_2$ in the nematic phase, its value would be approximately the same as in
      the isotropic phase.
\bibitem{footnote_faseprop}
    The multiple quantum spectra could also be displayed by using time proportional
      phase increment \cite{marion89}. Since we were not conditioned by the
      duration of the experiment we used another strategy to assure the frequency
      resolution needed.
\end{thebibliography}

\begin{appendix}
\section{Analysis of non-idealities in the spin-dynamics reversion sequence.} \label{ap:non-ideal}

    This appendix analyzes the effect of the non-secular terms of the dipolar Hamiltonian on the spin dynamics under the reversion sequence MREV8 (\fig{fig:Puls}(b)). The phase cycling of the component pulses aims to alternate the sign of the non-secular part between subsequent pulse pairs while keeping the same (inverted) sign of the secular part. Therefore, the average effect over the reversion and the free-evolution periods of this sequence is to cancel evolution under the secular dipolar Hamiltonian (see Chapter 2.\textbf{C}(b) from \cite{abragol82}).
    However, in actual non-ideal cases, there is a residual evolution due to non-secular terms. The time between pulses is usually set as small as possible (but not shorter than the pulses) to mitigate the residual effect. However, this strategy does not apply when long reversion periods are needed.

    We use an 8-spin system with the geometry and dipolar couplings of the $^1$H nuclei in a PAA$_{d6}$ molecule \cite{SegZam11}, and calculate the signals  corresponding to the experiment shown in \fig{fig:Puls}(a) and described in \sect{se:QEDetect}, under the MREV8 sequence.

    The left column of \fig{fig:CalcCM_PAAd6} shows the spectra corresponding to $\tau_1 = 5\un{\mu s}$ and the right column to  $\tau_1 = 20\un{\mu s}$.  The first row corresponds to $t_p = 0\un{\mu s}$, or equivalently to reverting the FID (single-spin single-quantum signal); rows (b) and  (c) correspond to preparation of the  $\mathcal{S}$-order, and $\mathcal{W}$-order, respectively, and in row (d) the preparation time of a compound state between \mbox{$\mathcal{S}$-} and $\mathcal{W}$-order . The total reversion time in the $\tau$ axis is determined by the number of blocks.

    Reversion is optimal when the characteristic time is small ($\tau_1 = 5\un{\mu s}$):  coherence amplitudes are practically independent of $\tau$ throughout the whole range ($1\un{ms}$), as seen in \fig{fig:CalcCM_PAAd6}(a1), (b1), (c1) and (d1). On the contrary, setting a much larger delay, $\tau_1 = 20\un{\mu s}$ introduces variations on the amplitudes of the different spectra with the reversion time $\tau$, as shown in \fig{fig:CalcCM_PAAd6}(a2), (b2), (c2) and (d2). Particularly, the amplitude of the zero-quantum coherence peaks in (b2) and (d2), respectively with $t_p = 47.5\un{\mu s}$ and $t_p = 195\un{\mu s}$ show a slow attenuation similar to the one observed in \figs{fig:ExpCM_5CB_S}(a) and (b), and \fig{fig:ExpCM_PAAd6}(a).
    Notice that when the initial state is prepared with $t_{p\mathcal{W}}$  as in  \fig{fig:CalcCM_PAAd6}(c2), zero-quantum coherence amplitude undulates with $\tau$.

    \insertfig{h}{fig:CalcCM_PAAd6}{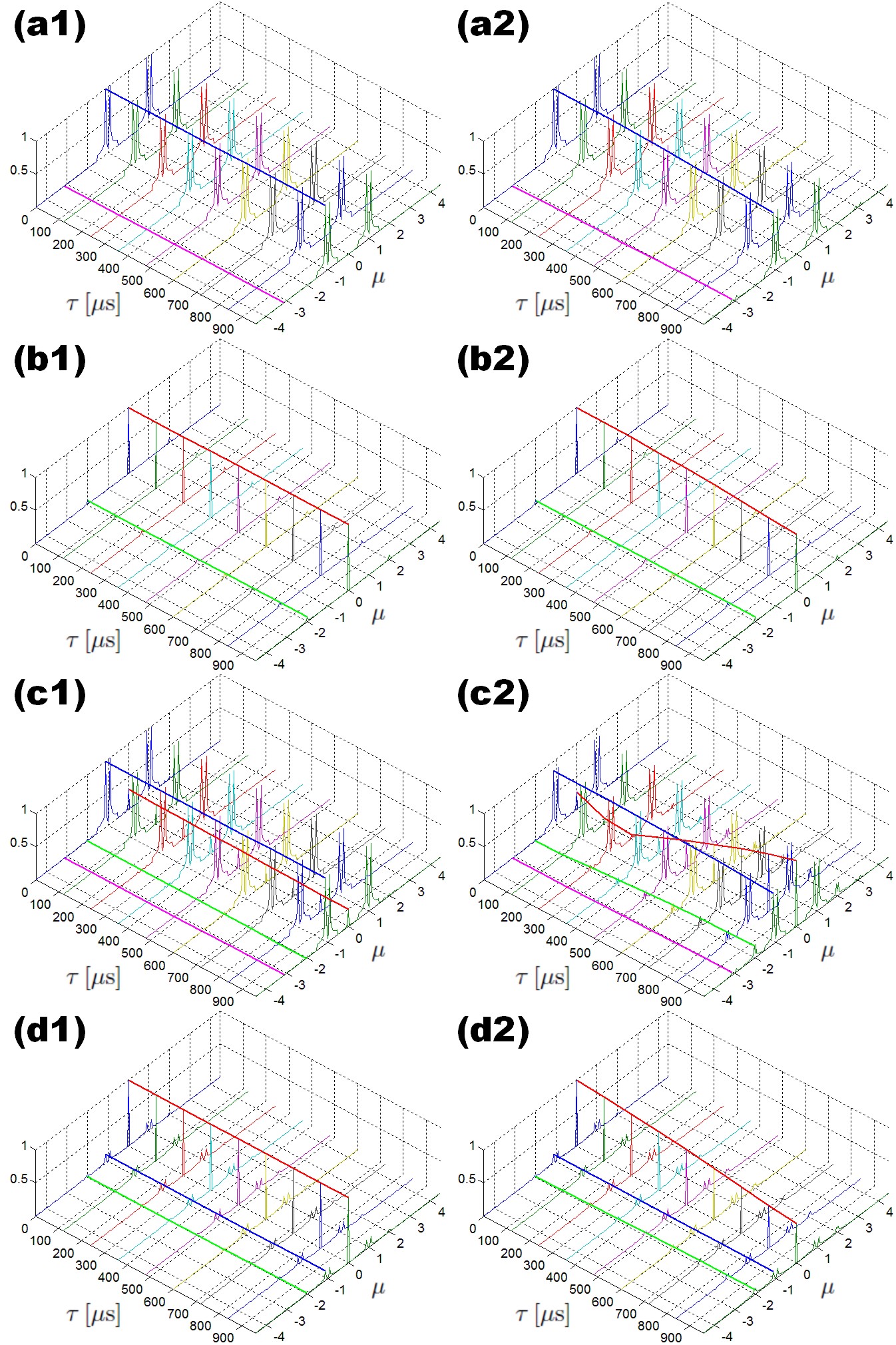}{9.0}{13.5}{Calculated output of the experiment in \fig{fig:Puls}(a) applied to a model spin system for the PAA$_{d6}$ molecule (dipole couplings from Ref.\cite{SegZam11}), under two versions of the MREV8 sequence: left column $\tau_1 = 5\un{\mu s}$; right column $\tau_1 = 20\un{\mu s}$. Row (a) $t_p = 0\un{\mu s}$, row (b) $t_p = 47.5\un{\mu s}$ ($\mathcal{S}$-order), row (c) $t_p = 92.5\un{\mu s}$  ($\mathcal{W}$-order), row (d) $t_p = 195\un{\mu s}$ (Compound state). As a guide for visualizing the evolution over $\tau$, the solid lines join the maximum peaks of different coherence spectra: $\mu=0$ red, $\mu=1$ blue, $\mu=2$ green, $\mu=3$ magenta.}

\end{appendix}

\end{document}